\documentclass[a4paper,11pt]{article}
\usepackage{graphicx}
\usepackage[space]{grffile}
\usepackage{latexsym}
\usepackage{textcomp}
\usepackage{longtable}
\usepackage{tabulary}
\usepackage{booktabs,array,multirow}
\usepackage{amsfonts,amsmath,amssymb}
\usepackage{natbib}
\usepackage{url}
\usepackage{hyperref}
\hypersetup{colorlinks=false,pdfborder={0 0 0}}
\usepackage{etoolbox}
\makeatletter
\usepackage{authblk}
\usepackage[utf8]{inputenc}
\usepackage[english]{babel}
\usepackage[top=1in, bottom=1in, left=1in, right=1in]{geometry}
\title{Moist Shallow Water Response to Tropical Forcing: Initial Value Problems}
\date{}
\author[1]{D.L. Suhas \thanks{Corresponding author: D.L. Suhas, suhasdl.mysore@gmail.com}}
\author[1,2]{Jai Sukhatme}
\affil[1]{Centre for Atmospheric and Oceanic Sciences, Indian Institute of Science, Bangalore, 560012, India}
\affil[2]{Divecha Centre for Climate Change, Indian Institute of Science, Bangalore, 560012, India}

\begin{document}
\maketitle

\begin{abstract}
The response of a spherical moist shallow water system to tropical imbalances in the presence of inhomogeneous saturation fields is examined. While the initial moist response is similar to the dry reference run, albeit with a reduced equivalent depth, the long time solution depends quite strikingly on the nature of the saturation field. For a saturation field that only depends on latitude, specifically, one with a peak at the equator and falls off meridionally in both hemispheres, height imbalances adjust to large-scale, low-frequency westward propagating modes. When the background saturation environment is also allowed to vary with longitude, in addition to a westward quadrupole, there is a distinct eastward propagating response at long times. The nature of this eastward propagating mode is well described by moist potential vorticity conservation and it consists of wave packets that arc out to the midlatitudes and return to the tropics and are predominantly rotational in character. 
In all the moist cases, initially formed Kelvin waves decay, and this appears to be tied to the mostly off-equatorial organization of moisture anomalies by rotational modes. Many of these basic features carry over to the response in the presence of realistic saturation fields derived from reanalysis based precipitable water. Considering an imbalance in the equatorial Indian Ocean, in the boreal summer, the long time eastward response is restricted to the northern hemisphere and takes the form of a wavetrain that passes over the Indian landmass into the subtropics, reaching across the Pacific to North America. In the boreal winter, the eastward mode consists of a subtropically confined rotational quadrupole along with the midlatitudinal disturbances. Thus, in addition to circumnavigating westward Rossby waves, slow eastward propagating modes appear to be a robust feature of the shallow water system with interactive moisture in the presence of saturation fields that vary with latitude and longitude.\\

\noindent \textit{Key Words : Initial Value Problem, Nonlinear Solutions, Moist Shallow Water Equations, Inhomogenous Saturation Background}

\end{abstract}

\section{Introduction}

The response of the atmosphere to near-equatorial forcing, usually in terms of a heat source, has been useful in developing an understanding of the waves that can exist at low latitudes \citep{matsuno1966}, as well as the non-axisymmetric nature of the tropical circulation \citep{gill1980}. Further, simplified dynamical models such as the shallow water system or single layer vorticity equations have proven to be useful in understanding a variety of phenomena such as emergent global teleconnections on the inclusion of base flows \citep{LauLim,Sardesh}, equatorial trapping \citep{zhangweb}, geostrophic adjustment of long-wave disturbances \citep{SRZ} and tropical stationary waves that form with continuous forcing \citep{Krau2007}.
Time dependent solutions in a linear shallow water setting \citep{MaTa,KaKh}, as well as in more complicated models, such as the linear and nonlinear primitive equations with episodic forcing restricted to specified timescales \citep{SG1,SG2}, steady forcing \citep{webster,JinHoskins} and the initial value problem \citep{HoskinsJin} with spatially varying background flows have also been pursued to examine the character of waves generated in the tropics, their propagation to extratropical regions and the evolution towards a steady state circulation. On a related note, the response to specific tropical phenomena such as the heating associated with Madden-Julian Oscillation \citep[MJO;][]{matt,BoHart,joy,kosovelj} and El Ni\~no \citep{Lin2007,shaman} have also been analysed to delineate their global character.

\noindent Quite reassuringly, on the steady front, reanalysis data suggests that at low latitudes, stationary waves in the upper troposphere are equatorial Rossby waves and are mainly a response to tropical heating and orography \citep{Dima}. 
Similarly, space-time spectral analysis of tropical fields (such as outgoing longwave radiation) that capture evolving disturbances show a signature of linear equatorial waves \citep{wk1999}. But, there are two caveats regarding the wavenumber-frequency spectra from data: (i) The wave speeds are significantly lower than predictions from the dry shallow water theory \citep{matsuno1966}. This reduction in speed is attributed to the coupling of moist processes with large scale dynamics resulting in the so called convectively coupled equatorial waves \citep[CCEWs;][]{Kiladis-rev}. (ii) More importantly, there is significant power in regions of the wavenumber-frequency diagram that do not correspond to traditional linear, dry equatorial waves --- specifically, the MJO \citep{Zhang}. Indeed, interactive moisture is believed to play a vital, possibly an essential role in the MJO \citep{sobel2013,adames2016}.

\noindent Thus, while small perturbations to the dry shallow water system produce the classical spectrum of tropical waves \citep{matsuno1966,MaTa,vallisbook}, realizing the importance of moist processes, attempts have been made to include the effect of moisture on dynamics while retaining the simplified nature of the shallow water framework. The intent of these simplified models is to examine if there are any new modes with moisture that do not exist in the dry system --- in particular, whether a signature of a MJO like system appears on the inclusion of moisture. In this regard, momentum equations with precipitation \citep{Yama}, or both condensation and bulk aerodynamic evaporative protocols and either an explicit \citep{RoZe1,RoZe2,Vallis}, or implicit 
\citep{Solod,Yang} treatment of water vapour and possibly even boundary layer convergence \citep{wang2016} have been explored. In fact, linear analyses that include moisture gradients have also been pursued \citep{Sobel,Sukhatme2,joyQG}. Indeed, all these systems show the presence of a slow, large-scale, eastward propagating mode, albeit via differing mechanisms, that is absent without moisture. Though it is worth noting that there is recent evidence for eastward propagating modons that develop under special initial conditions in a dry setting \citep{RoZe1}. Apart from MJO related investigations, moist dynamics in relatively simple models have also been utilized in understanding the genesis, structure and propagation of CCEWs \citep{kuang1,KhMa1,dias} and tropical cyclones \citep{schecter2009,LZ2016,RoZe4}. 

\noindent Here, following \citet{gill1982,Bouchut}, we consider the nonlinear dynamics of a layer of fluid that is subject to condensation and evaporation by means of mass loss and gain. Till date, as far as we are aware, there has been no systematic documentation of the nonlinear atmospheric response to tropical forcing in the presence of a spatially non-uniform background saturation field. In particular, we investigate the waves produced with saturation fields that range from being purely a function of latitude to those derived from seasonal mean maps of precipitable water in reanalysis data. Thus, akin to studies that elucidated the striking variations in the response to tropical heat sources in the presence of different mean flows \citep{LauLim,Sardesh}, we aim to catalogue the changing nature of the response with progressively more realistic background saturation fields. Section 2 contains the basic equations and a description of the initial conditions. Section 3 shows the response of a dry system and serves as a reference to compare the moist simulations. Section 4 considers the corresponding moist initial value problem in the presence of different background horizontal saturation fields. In particular, we proceed systematically with saturation fields that are purely functions of latitude, idealized forms that depend on latitude and longitude and finally saturation environments derived from reanalysis estimates of precipitable water. The manuscript then concludes with a discussion of results in Section 5.

\section{Equations and Numerical Setup}

The moist shallow water equations with mass and momentum forcing, radiative damping and momentum drag, precipitation and evaporation take the form \citep{gill1982,Bouchut},
\begin{align}
\frac{\partial \omega}{\partial t} + \nabla \!\cdot \!({\bf u} \omega_a) = - \frac{\omega}{\tau_m} + F_{\omega}, \nonumber \\
\frac{\partial \delta}{\partial t} - \textbf{k} \!\cdot \! \nabla \times (\textbf{u} \omega_a) = -\triangledown^2 (\frac{{\bf u} \!\cdot {\bf u}}{2} + gh) - \frac{\delta}{\tau_m} + F_{\delta} , \nonumber \\
\frac{\partial h}{\partial t} + \nabla \!\cdot \!({\bf u}h) =  -\frac{(h-H)}{\tau_r} + S - \chi \mathcal{(P - E)},  \nonumber \\
\frac{\partial q}{\partial t} + \nabla \!\cdot \!({\bf u}q) =  -\mathcal{P} + \mathcal{E}. 
\label{1}
\end{align}
Here, ${\bf u}=(u,v)$ is the horizontal flow, 
$\omega$ and $\omega_a$ are the relative and absolute vorticity, $\delta$ is the divergence and $h(x,y,t)$ is the depth of the fluid
($H$ is the mean undisturbed depth). The drag in the vorticity and divergence equations has the same timescale $\tau_m$; $\tau_r$ is the radiative damping timescale, $S$ is the mass forcing and $F_\omega,F_\delta$ are vorticity and divergence forcing terms. Further, $q$ is the column water vapour, $q_s$ is a prescribed moist saturation field, $\mathcal{P}$ and $\mathcal{E}$ are the moisture sink and source terms. $\chi$ is an effective specific heat for this system. The formulation of this system, nonlinear waves and fronts that develop in the presence of moisture, emergence of modon like solutions with special initial conditions as well as the adjustment of pressure anomalies on the equatorial $\beta$-plane with a uniform background saturation field have been explored in detail \citep{Bouchut,RoZe1,RoZe2}.

\noindent Precipitation is dependent on column water vapour \citep{MuEm}, and takes a Betts-Miller form \citep{Betts}, $\mathcal{P} = (q-q_s) \Theta (q-q_s)/ \tau_c$, where $\tau_c$ is the condensation timescale and $\Theta$ is 1 if $q>q_s$ and 0 otherwise.
Evaporation takes the similar form with $\mathcal{E} = (q_s-q)  \Theta (q_s-q)/ \tau_e$, where $\tau_e$ is the evaporation timescale. This form of $\mathcal{E}$ is suitable for low winds and has been used to demonstrate the enhancement of cyclonic vorticity in moist barotropic and baroclinic instabilities \citep{RoZe3}.  Therefore, a parcel either experiences condensation (if $q>q_s$) or evaporation (if $q<q_s$), and we set their timescales to be equal, i.e., $\tau_c=\tau_e$.  

\noindent Combining the last two equations in (\ref{1}), setting $m=h-\chi q$, we obtain,
\begin{equation}
    \frac{Dm}{Dt} + m \nabla \!\cdot \!{\bf u} = \frac{-(h-H)}{\tau_r} + S.
    \label{3}
\end{equation}
A moist potential vorticity (PV) equation that results is,
\begin{equation}
    \frac{D}{Dt}\bigg[ \frac{\omega_a}{m} \bigg] = \frac{\omega_a (h-H)}{m^2 \tau_r} - \frac{S \omega_a}{m^2} - \frac{\omega}{m \tau_m} +  \frac{F_{\omega}}{m}.
    \label{eqn:MoistPV}
\end{equation}
Thus, when the sources are switched off and $\tau_r,\tau_m \rightarrow \infty$, moist PV, i.e., ($\omega_a/m$) is materially conserved \citep{joyQG}. 

\noindent The shallow water system including the moisture equation is solved by a pseudo-spectral method in spherical geometry using the SHTns library \citep{shtns}. 
This code has been validated with a range of problems including the typical test case proposed by \citet{galewsky}. 
The use of spectral methods for moisture transport often leads to undershooting \citep{laprise, rasch}. This results in physically unrealistic negative moisture values, and is remedied by immediately resetting the negative values of $q$ to 0. Though, in our shallow water system, with small perturbations and fields being initially saturated along with the presence of evaporation, the problem is very rarely encountered.
All results are reported at a resolution of 512 (longitude) $\times$ 256 (latitude), triangularly truncated corresponding to a maximum resolved wavenumber of 170.
We use a 3$^{rd}$ order Adams-Bashforth integrator for time stepping. A $\triangle ^4$ hyperviscosity is used for small scale dissipation.
The mean depth of the fluid is fixed at 300 m so as to have phase speeds near that of the first baroclinic mode in the tropics \citep{Krau2007}. Given that our simulations cover the entire globe, it is possible to have a mean depth that varies with latitude \cite{joy}, but this introduces a mean flow that is known to affect the nature of the atmospheric response \cite{LauLim,Sardesh}. Thus, we choose to use a constant depth of 300 m which is appropriate for the tropical region. Planetary radius ($a$), rotation rate ($\Omega$) and acceleration due to gravity ($g$) are set to that of the Earth \citep{suhasQJ}. The timescales of both evaporation and condensation are relatively rapid, and are fixed to be 0.1 days (we have performed runs with varying $\tau_c,\tau_e$, and as long as these timescales are less than a day, they do not affect the nature of the results presented). In this work, we focus on initial value problems hence there is no large-scale drag or damping in the equations, specifically, $\tau_m,\tau_r = \infty$ in Equation \ref{1}. Steady and statistically stationary solutions, that include drag and damping, are explored in a companion paper. 
With regard to the choice for $\chi$ and $\max(q_s)$, this is based on the speed of moist Kelvin waves in the tropics. Specifically, with $H=300$ m, the dry Kelvin wave speed is $\approx 54$ m/s. Now, in the limit of rapid condensation, linear analysis of Equation \ref{1} suggests an equivalent depth of $h_e = H - \chi Q$. Setting $Q=\max(q_s)=0.05$, we choose $\chi=5000$ which yields a moist Kelvin wave speed of $\approx 22$ m/s. Another way to look up on this choice is to note 
that the moisture couples with $\chi$ to force the height in Equation \ref{1}, thus we set these two so that their product is less than, but of the same order as the mean depth of the fluid. 

\noindent We mainly focus on two kinds of initial conditions, (i) Gaussian height imbalance, here following \citet{gill1980}, an anomaly is introduced in the height field which is localized on the equator. This Gaussian anomaly is
of the form, $H_0  \exp[-((\phi - \phi_0) / \triangle \phi)^2] \ \exp[-((\lambda - \lambda_0) / \triangle \lambda)^2]$, 
where $H_0 = -10$ m is the peak magnitude, $\triangle \phi = 10^\circ$ and $\triangle \lambda = 30^\circ$ are the latitudinal and longitudinal half-width respectively, $\phi_0 = 0^\circ$ and $\lambda_0 = 180^\circ$ unless stated otherwise. A negative anomaly acts as a mass sink and is representative of lower tropospheric convergence. (ii) Random large-scale initial height field also localized to the tropical belt. 
The spectrum of this random perturbation takes the form, $\hat{A}e^{i\theta}$, 
where $\theta$ is a random number in $[0,2\pi)$ and $\hat{A}$ is the wavelength-dependent forcing amplitude with forcing restricted to total wavenumbers between 1 and 10. A meridional Gaussian profile (same as the previous case but is not a function of longitude) is multiplied to this random anomaly, thereby confining it to tropics but unbounded zonally. The longitudinally unbounded nature of this initial condition is motivated by  maps of wind divergence \citep{Hamil}, and rainfall that are spread throughout the tropics. In essence the two initial conditions consider the response to a localized coherent forcing (or heating) and scattered heating distributed throughout the tropics, respectively.

\section{Dry Response}

The response of a dry shallow water system to various tropical imbalances has been studied in the past; this includes its dependence on the initial scale of the disturbance \citep{MaTa}, nonlinear geostrophic adjustment of long-wave perturbations \citep{SRZ}, the differing nature of equatorial adjustment with the strength and aspect ratio of initial pressure anomalies \citep{RoZe2} and the wavenumber-frequency characteristics of freely evolving small scale random initial conditions \citep{vallisbook}. Here we document the waves that emerge for ready comparison with moist simulations, and also touch on some novel aspects such as the rotational-divergent makeup of the response \citep{Yuan, salmonbook}, spatial spectra \citep{vzagar}, and its eastward and westward propagating components. 

\begin{figure}
    \centering
    \includegraphics[width=\textwidth]{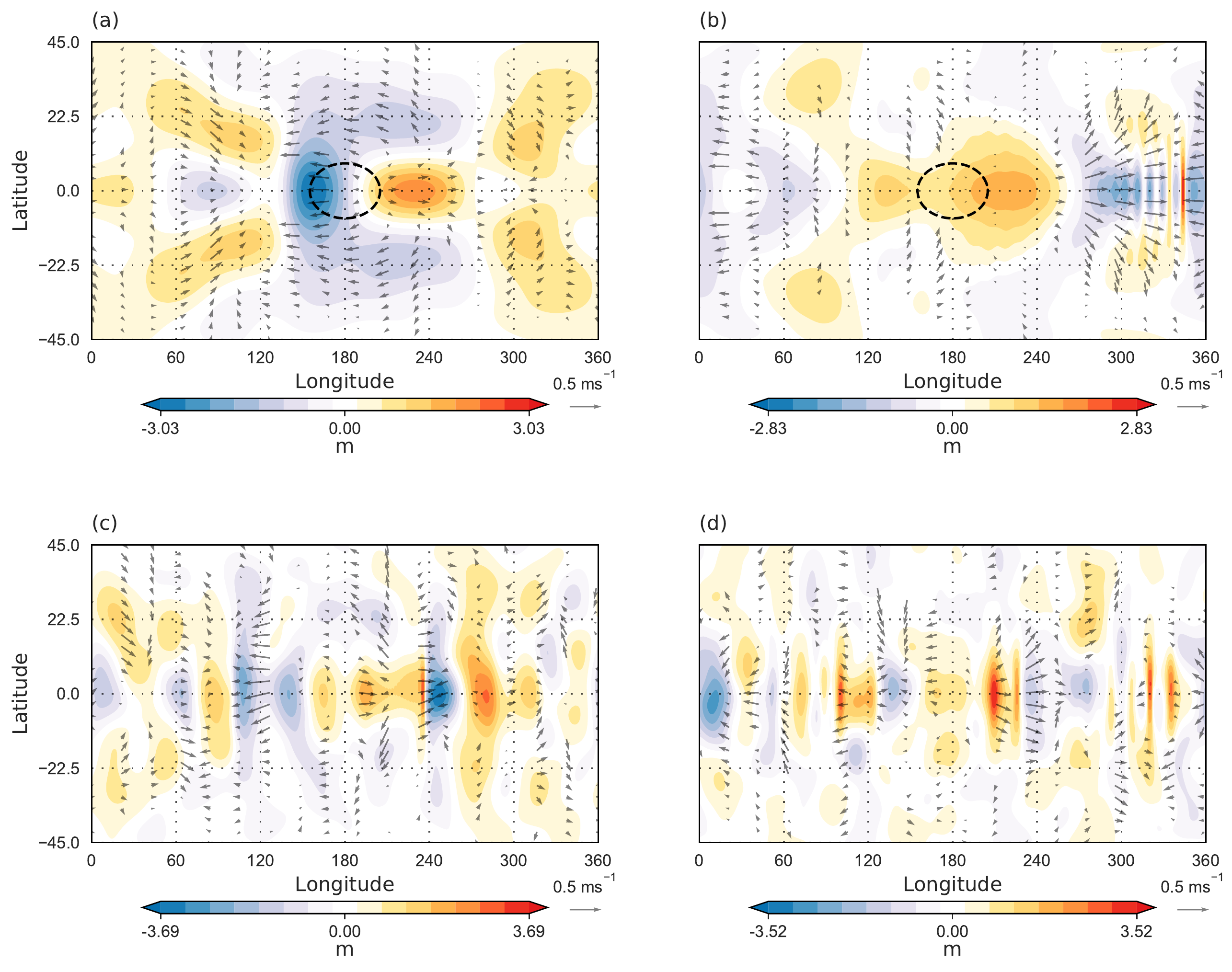}
    \caption{Height anomaly with velocity anomaly quivers for the dry case. Panels (a) and (b) are the response to a Gaussian mass sink (indicated by a black circle) at Day 25 and 250, respectively. The lower two panels (c) and (d)  are for a large-scale tropical random height field imbalance at Day 25 and 250.}
    \label{fig:gillDryHeight}
\end{figure}

\begin{figure}
    \centering
    \includegraphics[width=\textwidth]{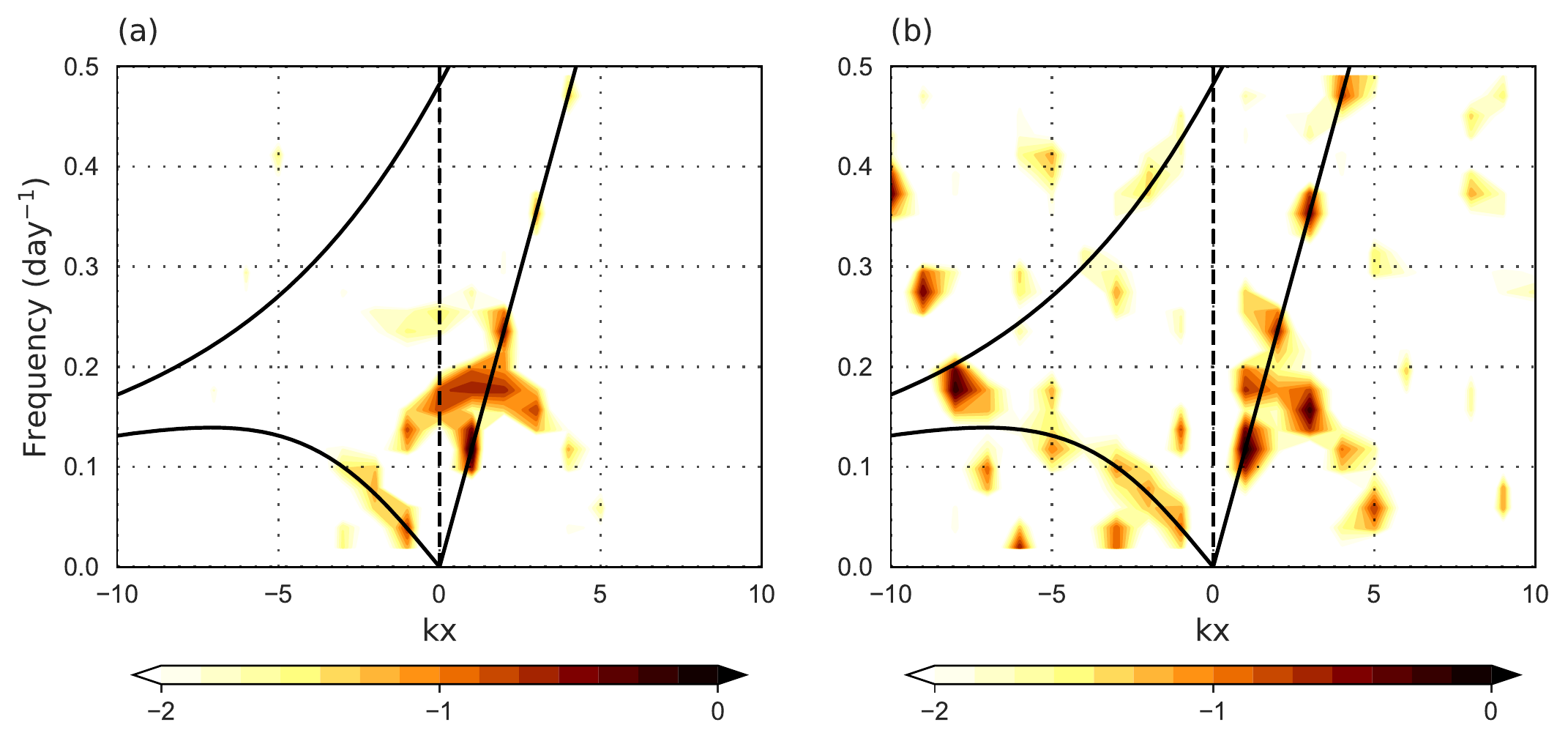}
    \caption{Wavenumber-frequency plot of height anomaly field for the dry case (a) Gaussian and (b) random initial perturbations. A 50 day period (Day 251-300) was used to construct the diagram. The latitudinal band used is from 15$^\circ$N to 15$^\circ$S. The power is normalized and plotted in log scale. Dispersion curves are plotted for an equivalent height of 300 m.}
    \label{fig:gillDryWK}
\end{figure}

\noindent Starting with a height imbalance, the evolution of height and velocity anomalies for the dry problem are shown in Figure \ref{fig:gillDryHeight}. Specifically, responses to a Gaussian mass sink and a large-scale random height perturbation are shown in the upper and lower panels, respectively. In both cases, adjacent highs and lows are evident on the equator and these spread out into off-equatorial gyres. On comparing the upper and lower panels of Figure \ref{fig:gillDryHeight}, the initial random height forcing appears to generate a spectrum of large to small scale features, whereas the response to a Gaussian forcing remains restricted to fairly large spatial scales. A wavenumber-frequency analysis of the solutions is shown in Figure \ref{fig:gillDryWK}. In each case, this diagram is constructed using a window of 50 days (Days 251-300) of the simulation. Note that a Hanning window was used to reduce the spectral leakage. Interestingly, late and early time (not shown) diagrams do not show much change, i.e., the response generated lasts through the course of the entire simulation (verified till 1000 days). Quite clearly, as in linear simulations \citep{MaTa}, the system generates a family of Rossby and Kelvin waves with an equivalent depth of about 300 m (which was the mean depth of the fluid used in the simulation). Further, the equatorial deformation radius ($\sqrt{{(g h_e)}^{1/2}/\beta}$) is about 14$^\circ$, and thus the response spreads out to a significantly larger latitudinal extent than this length scale. Noticeably, the eastward Kelvin peaks have more power and smaller time periods than the westward Rossby waves. Specifically, the eastward (westward) components have periods of about 8-9 days (25-28 days), respectively.
While the Rossby and Kelvin portion of the diagrams are similar in the two cases, the random height initial condition appears ``busier" with activity on the mixed Rossby-Gravity branch as well as isolated peaks scattered through the wavenumber-frequency domain. But by and large, even though this is a fully nonlinear solution, the wavenumber-frequency characteristics of the solution lie along the linear dispersion curves \citep[as can also be seen in,][]{vallisbook} . This remarkable feature is reminiscent of equatorially trapped solutions in the nonlinear non-divergent barotropic system \citep{Const} and has been noted in nonlinear geostrophic adjustment on an equatorial $\beta$-plane; in particular, long-wave disturbances were observed to evolve into slow Rossby and Kelvin waves that were split from the fast inertia-gravity wave family \citep{SRZ}.

\begin{figure}
    \centering
    \includegraphics[width=\textwidth]{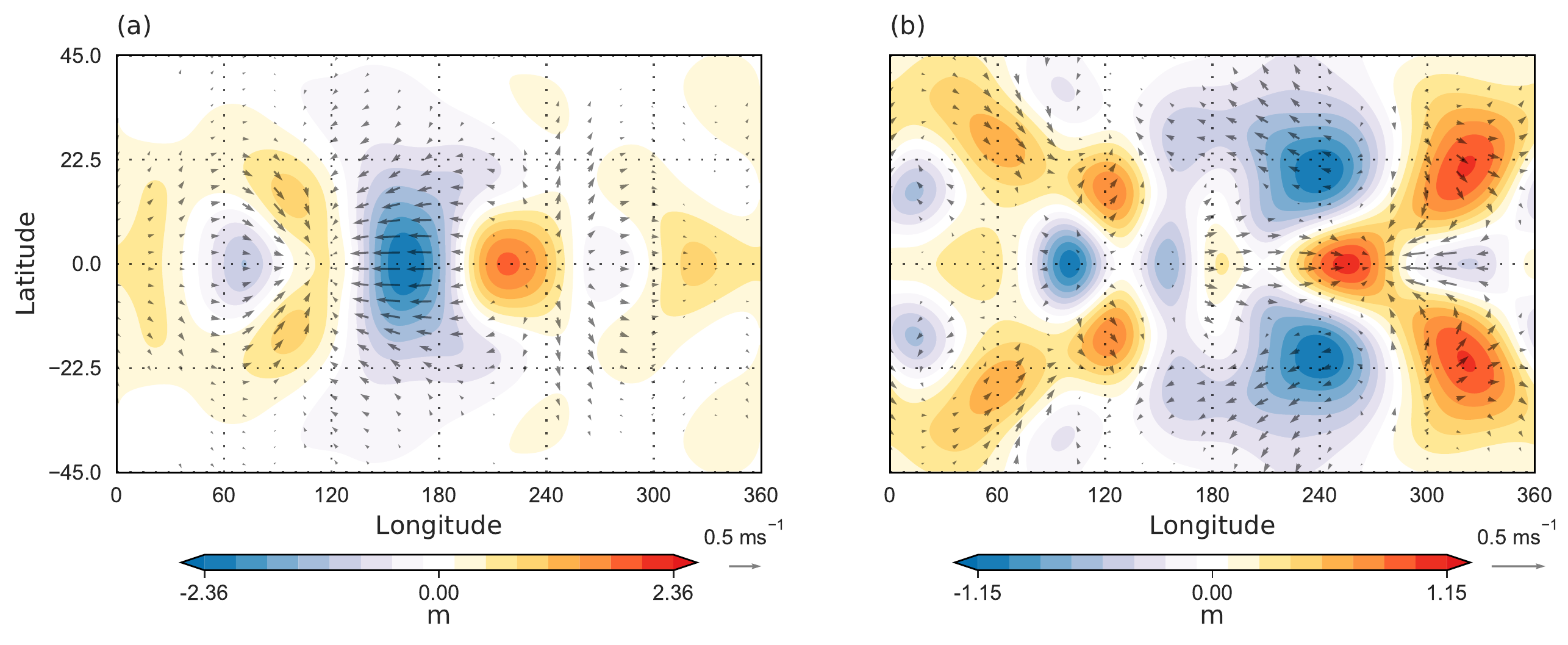}
    \caption{(a) Eastward and (b) westward component of height anomaly at Day 25 for the dry simulation. These are for the Gaussian height imbalance. First ten zonal wavenumbers are retained and a 50 day window is used to extract the eastward and westward components.}
    \label{fig:gillDryFiltHeight}
\end{figure}

\begin{figure}
    \centering
    \includegraphics[width=\textwidth]{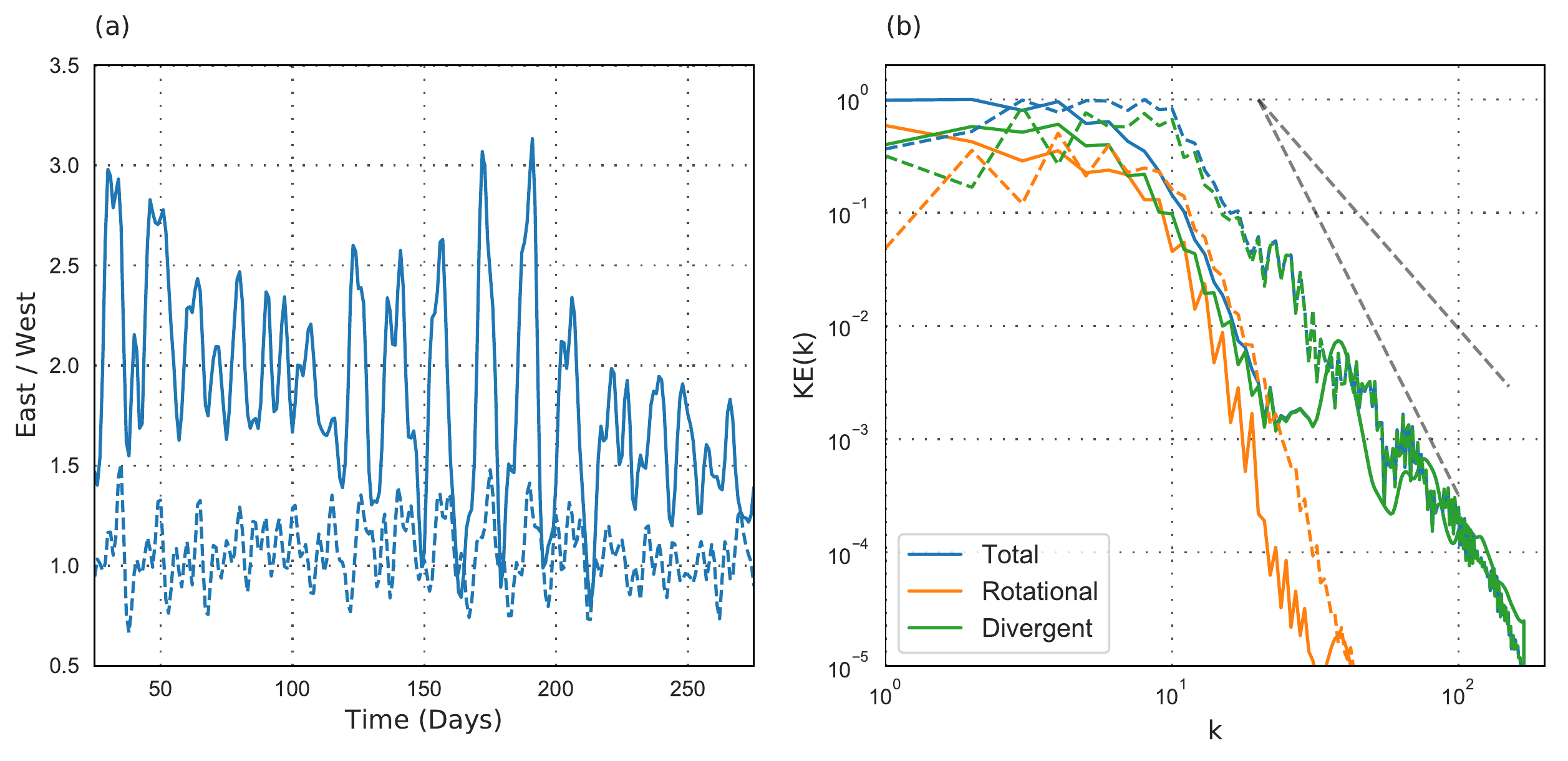}
    \caption{(a) Evolution of the ratio of eastward to westward component of Energy (KE+PE) with time for the dry case. (b) KE spectra averaged over Days 200-300. Grey dashed lines in (b) have slopes of -3 and -5. In both panels, solid lines: Gaussian and dashed lines: random initial perturbations.}
    \label{fig:gillDryEnergyTimeSpectra}
\end{figure}

\noindent Based on the wavenumber-frequency diagrams, a decomposition of the solution into eastward and westward moving structures is shown in Figure \ref{fig:gillDryFiltHeight}. Here, we have used a window of 50 days to reconstruct the eastward and westward signals and we show these components for the Gaussian initial condition.
The decomposition is robust with respect to changes in the window length. As we would expect, this yields off-equatorial rotational Rossby gyres (westward), and Kelvin waves (eastward) consisting of high-low pairs with maximum amplitude on the equator that travel in the eastward direction. Further, Rossby waves of different scales are visible in the westward propagating component of the solution. The contribution of the eastward and westward components to the kinetic and potential energy as a function of time is shown in the first panel of Figure \ref{fig:gillDryEnergyTimeSpectra}. As expected from the wavenumber-frequency diagram, these contributions fluctuate with time and moreover, the eastward component has greater energy than the westward portion. This is especially clear for the Gaussian case, while the ratio of the east to west energy hovers around unity for the random initial condition.

\noindent A spatial power spectrum of the kinetic energy (KE), shown in the second panel of Figure \ref{fig:gillDryEnergyTimeSpectra}, suggests the energy spreading out from the initially large scale localized source. For the Gaussian initial condition, there are some signs of power-law scaling (with a -5 exponent), but the range is quite limited. The spectrum for the random height imbalance suggests a much more robust transfer of energy to small scales and a power-law scaling appears more clear (with a -3 exponent). The contribution of rotational portion to the KE is comparable to the divergent part at the largest scales, but is quite small at smaller scales. A map of the rotational and divergent components of the flow (not shown) aligns quite well with the westward and eastward response, respectively. These results from the dry system serve as a ready reference for comparison with the moist simulations that are the main focus of this work\footnote{Other explorations of the dry system, especially the relative roles of divergent and rotational components can be found in the Masters project report of A. Baksi \citep{baksi}.}.

\section{Moist Response}
\subsection{Moist solutions with a purely latitudinal saturation profile}
\begin{figure}
    \centering
    \includegraphics[width=\textwidth]{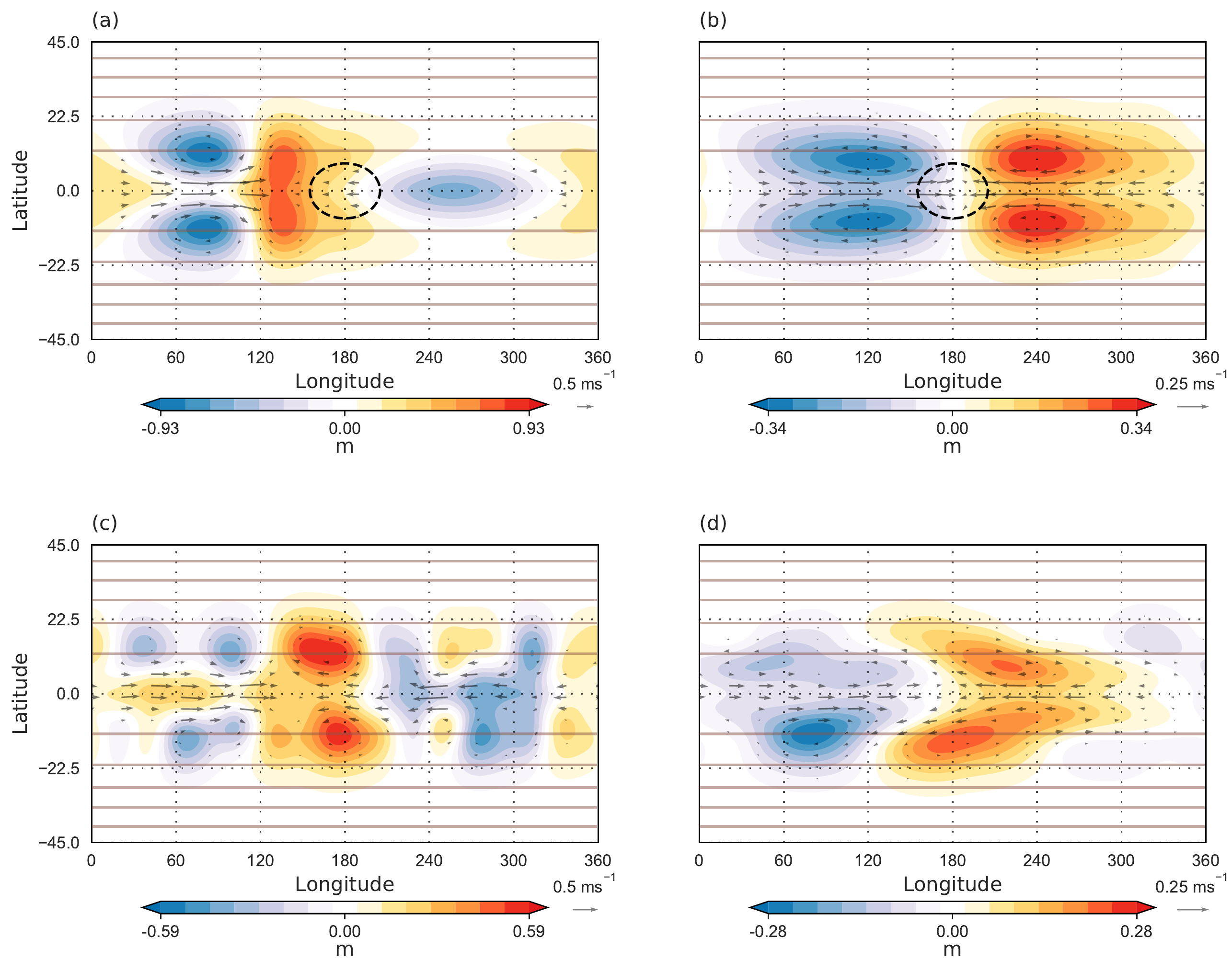}
    \caption{Height anomaly with velocity anomaly quivers when $q_s$ is a function of latitude with a maximum at the equator (shown in brown contours). The saturation contours are in intervals of 0.004 m with the one near the equator having a magnitude of 0.048 m. 
    Panels (a) and (b) are the response to a Gaussian mass sink (indicated by a black circle) at Day 25 and 250. Panels (c) and (d) are for a large-scale random height field imbalance at Day 25 and 250.}
    \label{fig:gillQsatLatHeight}
\end{figure}

\begin{figure}
    \centering
    \includegraphics[width=\textwidth]{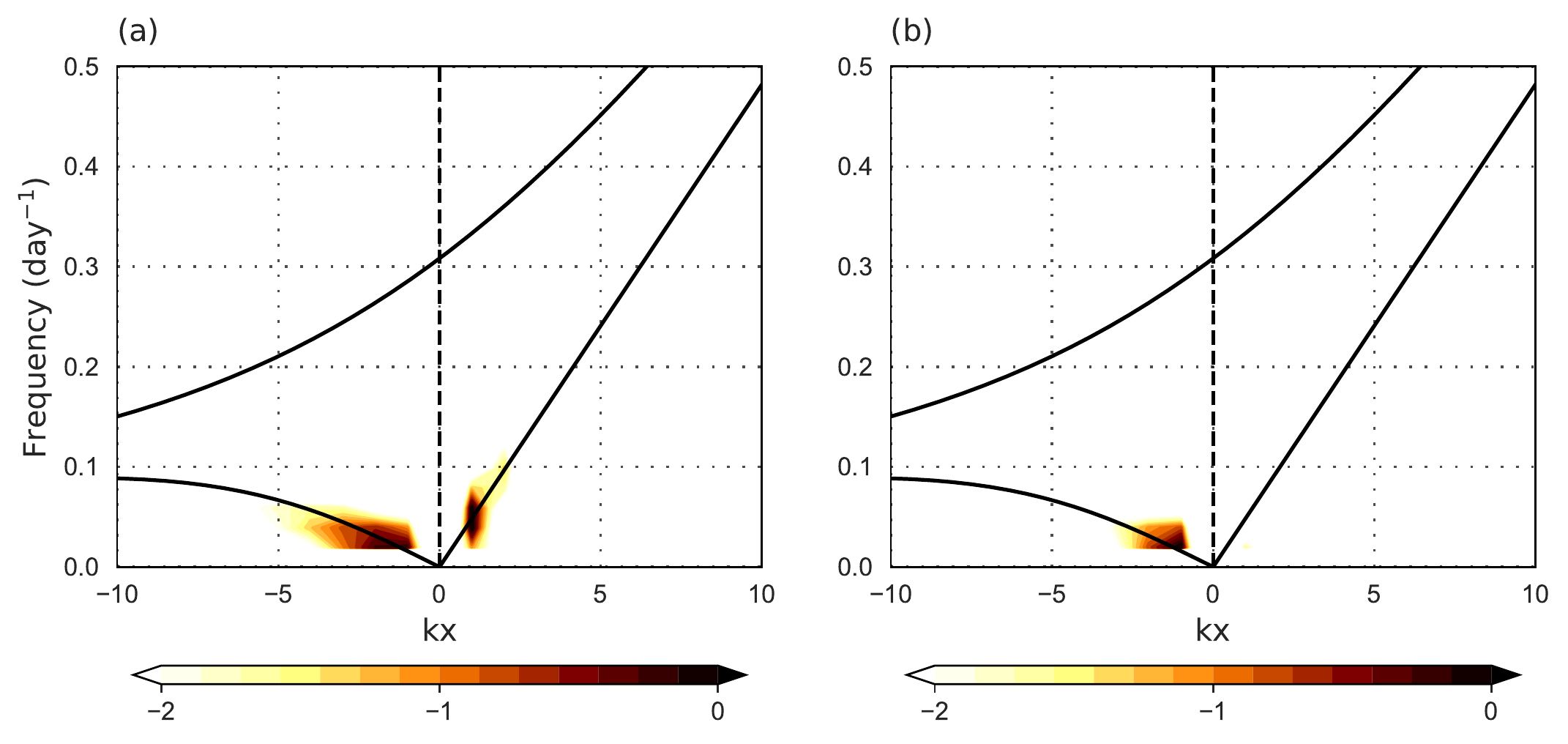}
    \caption{Wavenumber frequency plot of height anomaly field for (a) early (Day 1 - 50) and (b) late (Day 251-300) periods, when $q_s$ is a function of latitude with a maximum at the equator. The system is initially forced by a Gaussian mass sink. The latitudinal band used is from 15$^\circ$N to 15$^\circ$S. The power is normalized and plotted in log scale. Dispersion curves are plotted for an equivalent depth of 50 m.}
    \label{fig:gillQsatLatWK}
\end{figure}

The corresponding solutions for the moist simulation are now examined. Here, we first consider the case where the saturation field is purely a function of latitude.
In particular, $q_s$ has a maximum at the equator and falls off with increasing latitude. This idealized form of $q_s$ is motivated by a host of advection-condensation and aquaplanet studies \citep[see, for example,][]{Ray,GS2006,SY,blackburn}. With this form of $q_s$, Figure \ref{fig:gillQsatLatHeight} shows the height and velocity anomalies of the atmospheric response, where the upper two panels are for a Gaussian sink and the lower two for a random height imbalance.  Figure \ref{fig:gillQsatLatWK} shows the early and late time wavenumber-frequency diagrams for the Gaussian case (the random height imbalance produces a similar response and is not shown). Early in the evolution (before Day 50), we see Rossby and Kelvin waves, though with a much smaller equivalent depth (about 50 m). The smaller equivalent depth can be estimated by a linear analysis of Equation \ref{1} in the rapid evaporation-condensation limit, this yields $h_e = H - \chi Q$, where $H$ is the mean depth of the fluid and $Q$ is the scale factor for the precipitable water. As the wavenumber-frequency diagrams are constructed from data within 15$^\circ$N/S, the domain averaged $q_s \approx \max(q_s)$, i.e., we are probably justified in taking $Q=0.05$. With $\chi=5000$, this yields $h_e=50$ m.

\noindent Noticeably, comparing the first panels of Figures \ref{fig:gillDryWK} and \ref{fig:gillQsatLatWK}, even at early times, smaller time period disturbances are absent in the moist case when compared to the dry simulation. This lack of small time period activity appears to be a manifestation of the tendency of convective coupling to slow down the evolution of the system and reduce the speeds of propagation of equatorial waves \citep{Kiladis-rev}. In fact, given a smaller deformation radius (approximately 9$^\circ$) these solutions are confined to the subtropics with almost no signature extending out into the midlatitudes. Further, as seen in Figure \ref{fig:gillQsatLatWK}, at late times, energy is concentrated in large-scale Rossby waves (westward) with almost no eastward propagating features. 
Thus, the initial height imbalance adjusts to a large scale, westward propagating Rossby quardupole\footnote{This adjustment to mainly westward propagating waves is also noted in the moist system (with the saturation field being only a function of latitude) when the initial imbalance is in the vorticity field, or is purely divergent in nature.}.

\begin{figure}
    \centering
    \includegraphics[width=\textwidth]{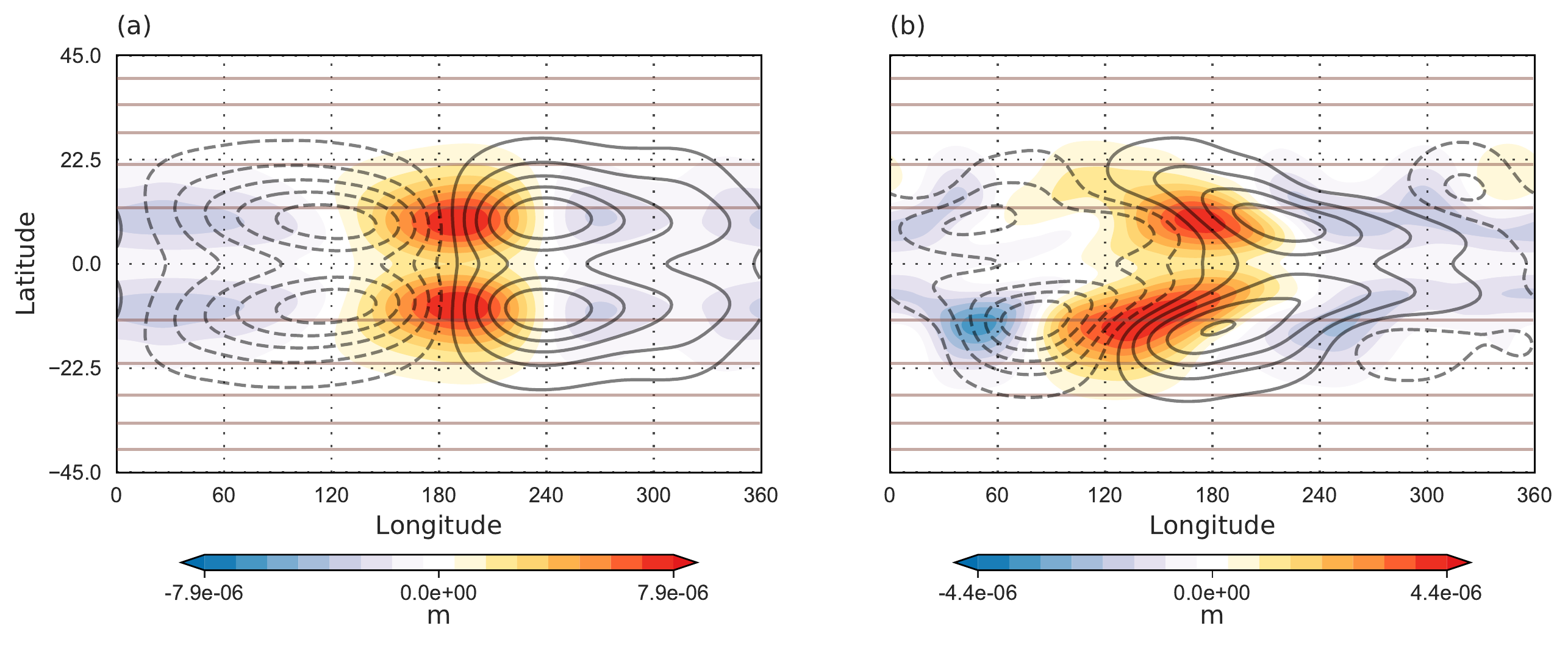}
    \caption{Moisture anomaly (deviation from the saturation state) at Day 250 for the (a)  Gaussian sink and (b) random height imbalance. Positive (negative) values indicate the region of precipitation (evaporation). Height anomalies are displayed in black contours. Here, $q_s$ is a function of latitude with a maximum at the equator (shown in brown contours). The saturation contours are in intervals of 0.004 m with the one near the equator having a magnitude of 0.048 m.}
    \label{fig:gillQsatLatQp}
\end{figure}

\noindent The time period of the westward moving quadrupole is approximately 75 days, which is also noticeably longer than its dry westward component. Note that, as our WK plots use a 50 day window, we use Hovm\"oller plots (not shown) to  estimate the time period.
The evaporation and condensation fields that go along with this emergent structure, in the Gaussian and random cases, are shown in Figure \ref{fig:gillQsatLatQp}. In both cases, condensation (evaporation) occurs to the west (east) of the anticyclonic gyres. This is consistent with moist (dry) air experiencing condensation (evaporation) as it moves anti-cyclonically away from (towards) the equator. Further, moist interactions, i.e., condensation and evaporation, aligns well with the regions of convergence and divergence, respectively. Thus, even though the initially imposed convergence is on the equator, at late times this field attains a maximum off the equator as dictated by the flanks of the adjoining Rossby gyres. 

\begin{figure}
    \centering
    \includegraphics[width=\textwidth]{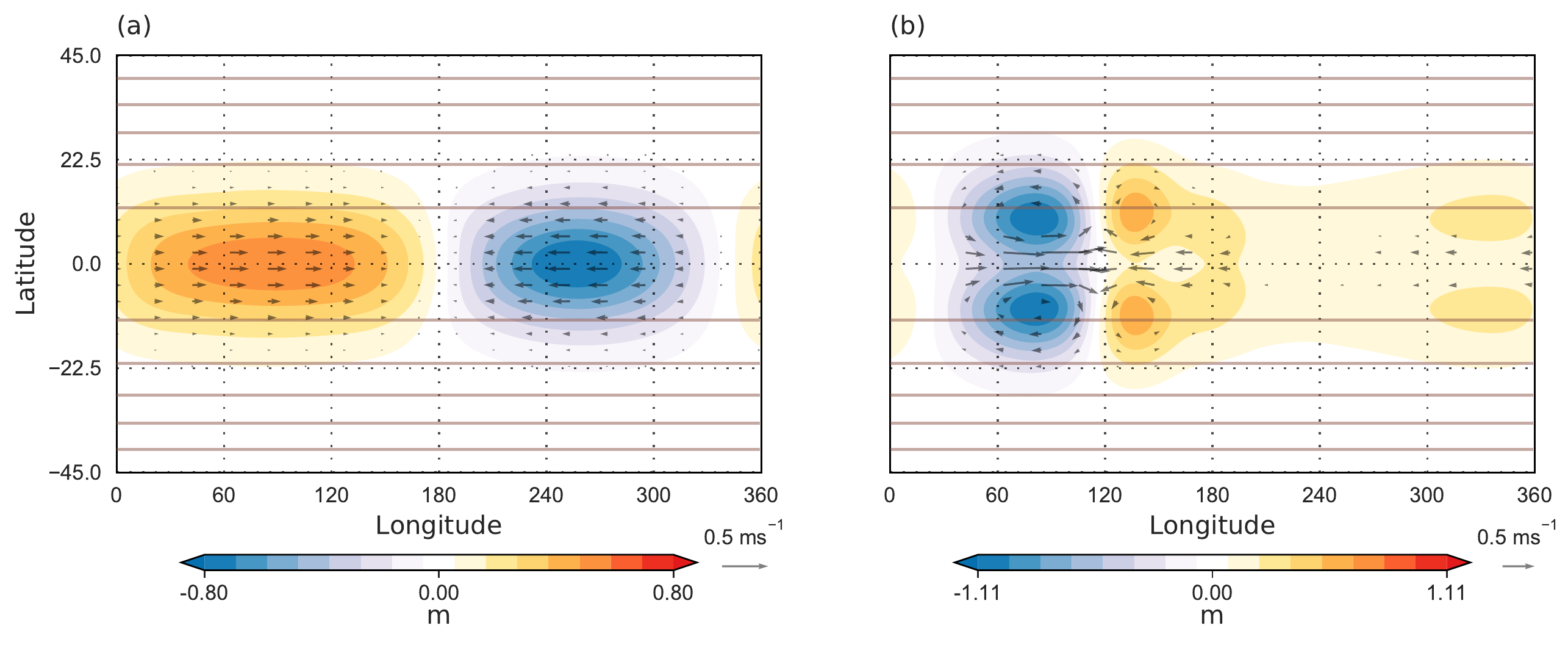}
    \caption{(a) Eastward and (b) westward component of height anomaly for the moist case at Day 25 when $q_s$ is a function of latitude with a maximum at the equator (shown in brown contours). First ten zonal wavenumbers are retained and a 50 day window centred at Day 25 is used. The system is initially perturbed by a Gaussian mass sink as shown in Figure \ref{fig:gillQsatLatHeight}.}
    \label{fig:gillQsatLatFiltHeight}
\end{figure}

\begin{figure}
    \centering
    \includegraphics[width=\textwidth]{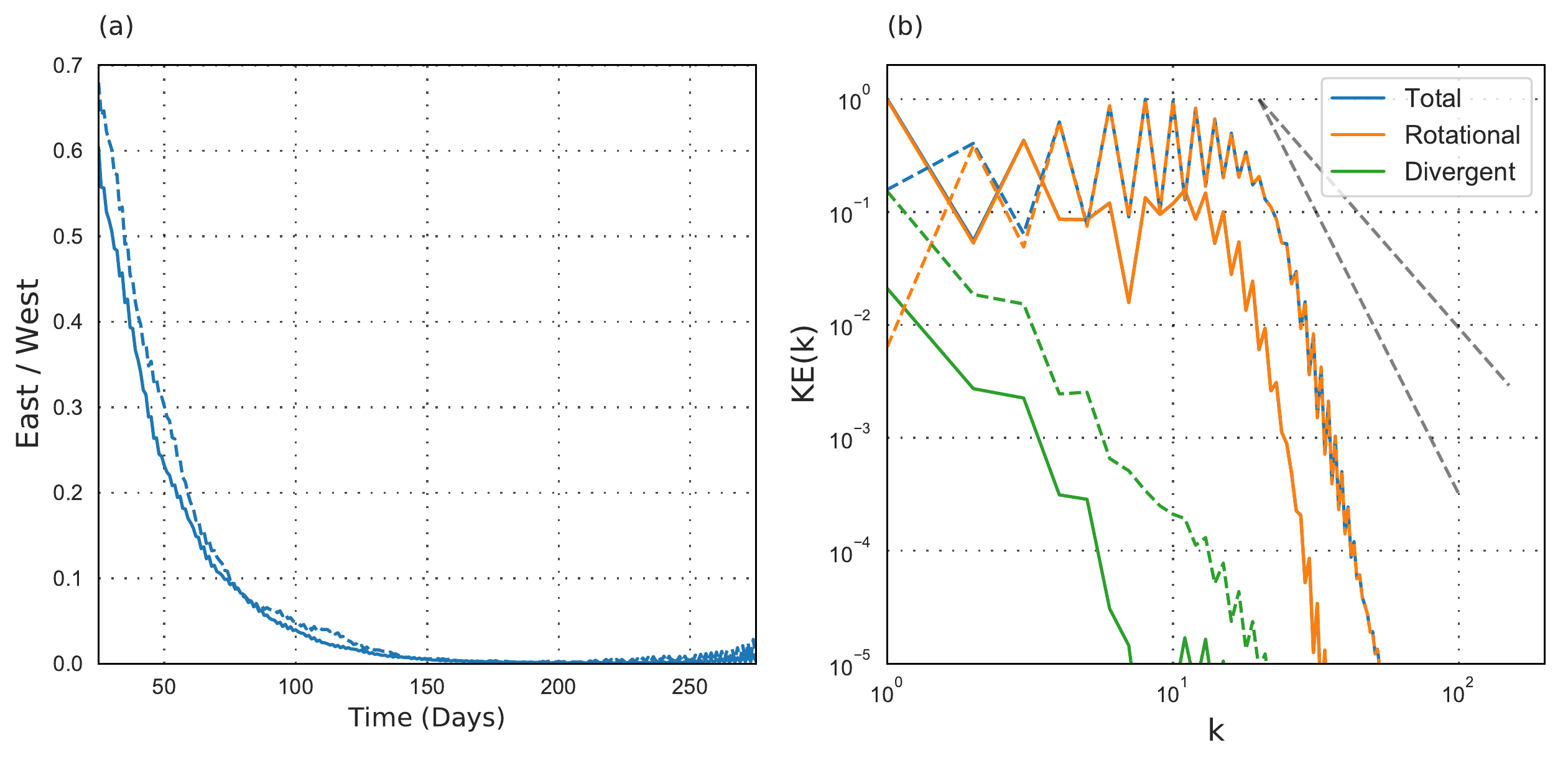}
    \caption{(a) Evolution of eastward to westward ratio of energy (KE+PE) with time and (b) KE spectra (averaged over Days 200-300). Here, $q_s$ is a function of latitude with a maximum at the equator. Grey dashed lines in (b) have slopes of -3 and -5. In both panels, solid lines are for the Gaussian sink and dashed lines are for the random height imbalance.}
    \label{fig:gillQsatLatEnergyTimeSpectra}
\end{figure}

\noindent A decomposition of the solution into eastward and westward portions for the Gaussian case at early times is shown in Figure \ref{fig:gillQsatLatFiltHeight}. A clear large-scale Kelvin wave is seen in the eastward direction while Rossby gyres appear in the westward component of the solution. As time goes on, the eastward component dies out while the Rossby gyres consolidate into a well formed quadrupole structure (Day 250 in Figure \ref{fig:gillQsatLatHeight}). The loss of the eastward features is seen via the decay of the  eastward to westward energy in Figure \ref{fig:gillQsatLatEnergyTimeSpectra}. The decay of an initially formed Kelvin wave during moist adjustment can also be seen in experiments with constant background saturation \citep{RoZe2}. The reason for this is not entirely clear, but as seen in Figure \ref{fig:gillQsatLatQp}, and as noted by \citet{adames2016}, moistening by anomalous winds is largely dominated by Rossby waves and is not conducive for Kelvin wave height anomalies. In particular, akin to the broader discussion by \citet{Frier} on the lack of Kelvin wave activity in idealized general circulation models with long convective adjustment timescales, the convergence region associated with the Kelvin waves near the equator no longer aligns with the off-equatorial moisture anomalies.
The spatial KE spectra from Day 200-300 suggest that most of the energy is in large length scales. Indeed, this is due to the fact that, in contrast to the dry run, by this time, almost all the energy in the moist simulation is contained in the rotational (westward propagating) portion of the flow and this is not transferred to smaller scales \citep[see, for example,][]{Yuan}. 

\subsection{Moist solutions with latitudinal and longitudinal saturation profile}

\begin{figure}
    \centering
    \includegraphics[width=\textwidth]{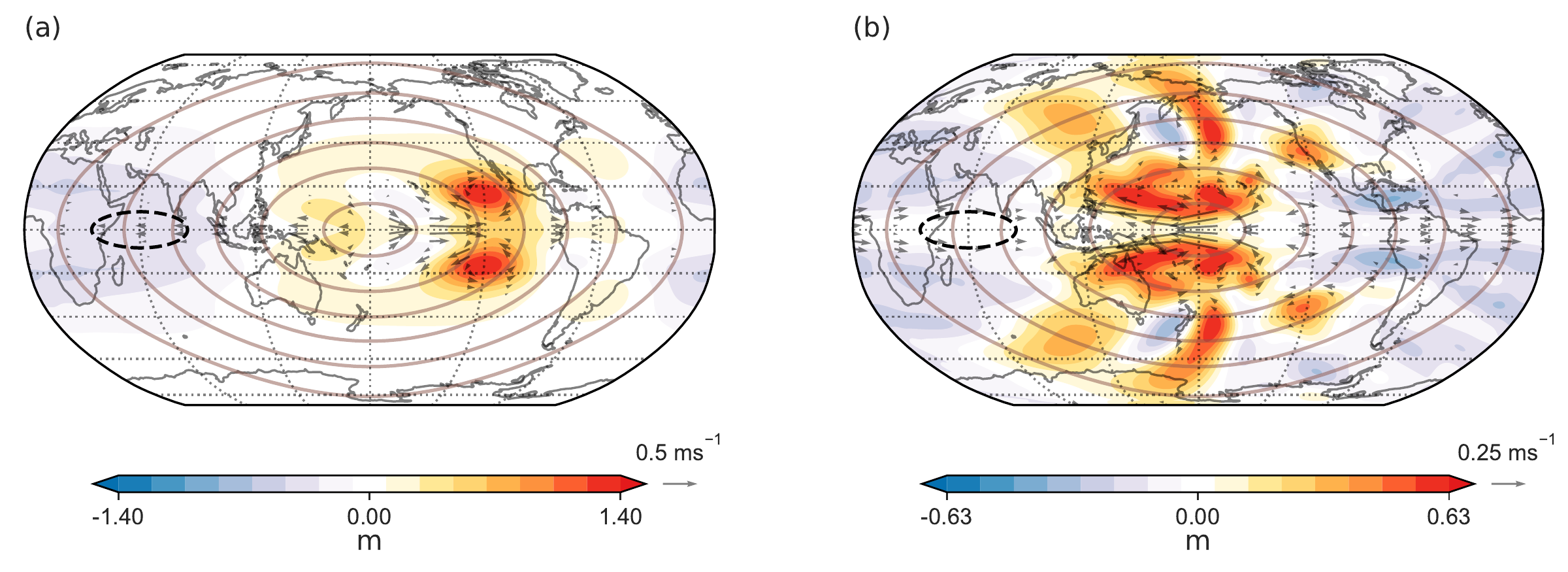}
    \caption{Height anomaly with velocity anomaly quivers at (a) Day 25 and (b) Day 250. $q_s$ is a function of latitude and longitude with a maximum at the equator and 180$^\circ$ longitude (shown in brown contours). The saturation contours are in intervals of 0.008 m, with zero contour not shown.The system is initially forced by a Gaussian mass sink (indicated by a black circle) to the west of the peak in $q_s$.}
    \label{fig:gillQsatBothHeight}
\end{figure}

In reality, precipitable water in the troposphere has marked variations in the latitudinal and longitudinal directions \citep{Sukhatme1,joyQG}. Taking this into account, we now examine the response to an initial height imbalance with $q_s$ decaying in both the directions with its peak centred at the equator and 180$^\circ$ longitude. Here, the position of the source of initial height imbalance can be important, specifically, we have three cases where the initial source is to the west, aligned with or to the east of the peak in $q_s$. It turns out that the position of the source affects the response, but only in the initial stages. Indeed, depending on where the source is located, the Rossby wave response is slow, muted and fast, when the source is to the east of, aligned with and to the west of the peak in $q_s$, respectively. Here, we present results for the case when the source is to the west of the peak in $q_s$. The height and velocity anomalies for a Gaussian imbalance are shown in Figure \ref{fig:gillQsatBothHeight} (the random case produces a similar response and is not shown). As a guide to the eye, a map is included beneath the height anomalies. Thus, the initial forcing is located in the equatorial Indian Ocean. A striking feature of the solution is the presence, at later times (Day 250 in Figure \ref{fig:gillQsatBothHeight}), of a wavetrain that expands out to the midlatitudes. A tropical wavenumber-frequency diagram (not shown) indicates the presence of Rossby and Kelvin waves at early times and follows linear dispersion curves with an equivalent depth of 160 m. This matches quite well with $h_e = H - \chi Q$, where $Q$ is the tropical (15$^\circ$ band straddling the equator) average of $q_s$. 

\begin{figure}
    \centering
    \includegraphics[width=0.5\textwidth]{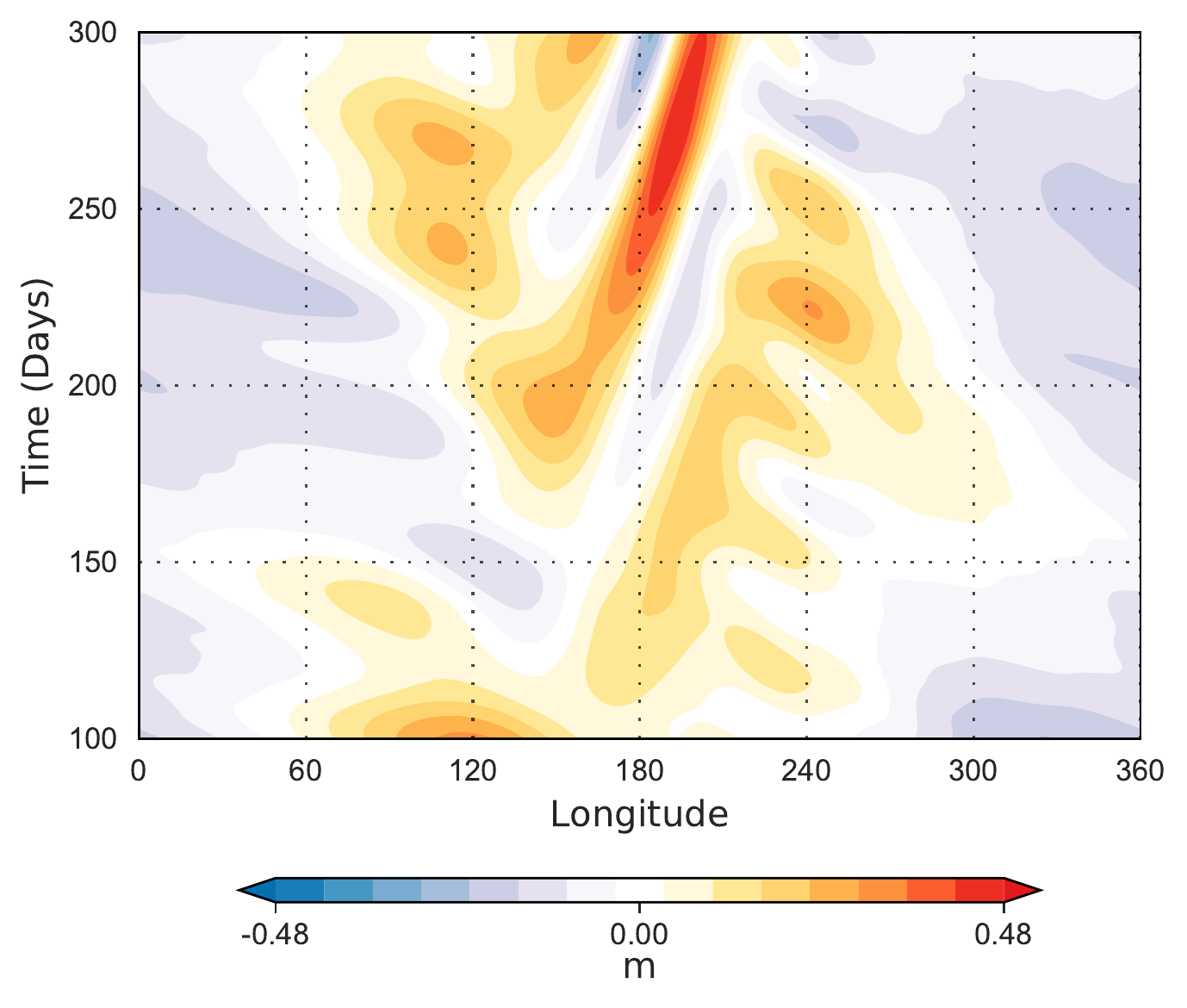}
    \caption{Ho{\"v}moller plot of height anomalies for the moist case where $q_s$ is a function of latitude and longitude with a maximum at the equator and 180$^\circ$ longitude. The system is initially forced by a Gaussian mass sink to the west of the peak in $q_s$ as shown in Figure \ref{fig:gillQsatBothHeight}. The height anomalies are averaged over the latitudes 30$^\circ$N to 60$^\circ$N. }
    \label{fig:gillQsatBothHovm}
\end{figure}

\begin{figure}
    \centering
    \includegraphics[width=\textwidth]{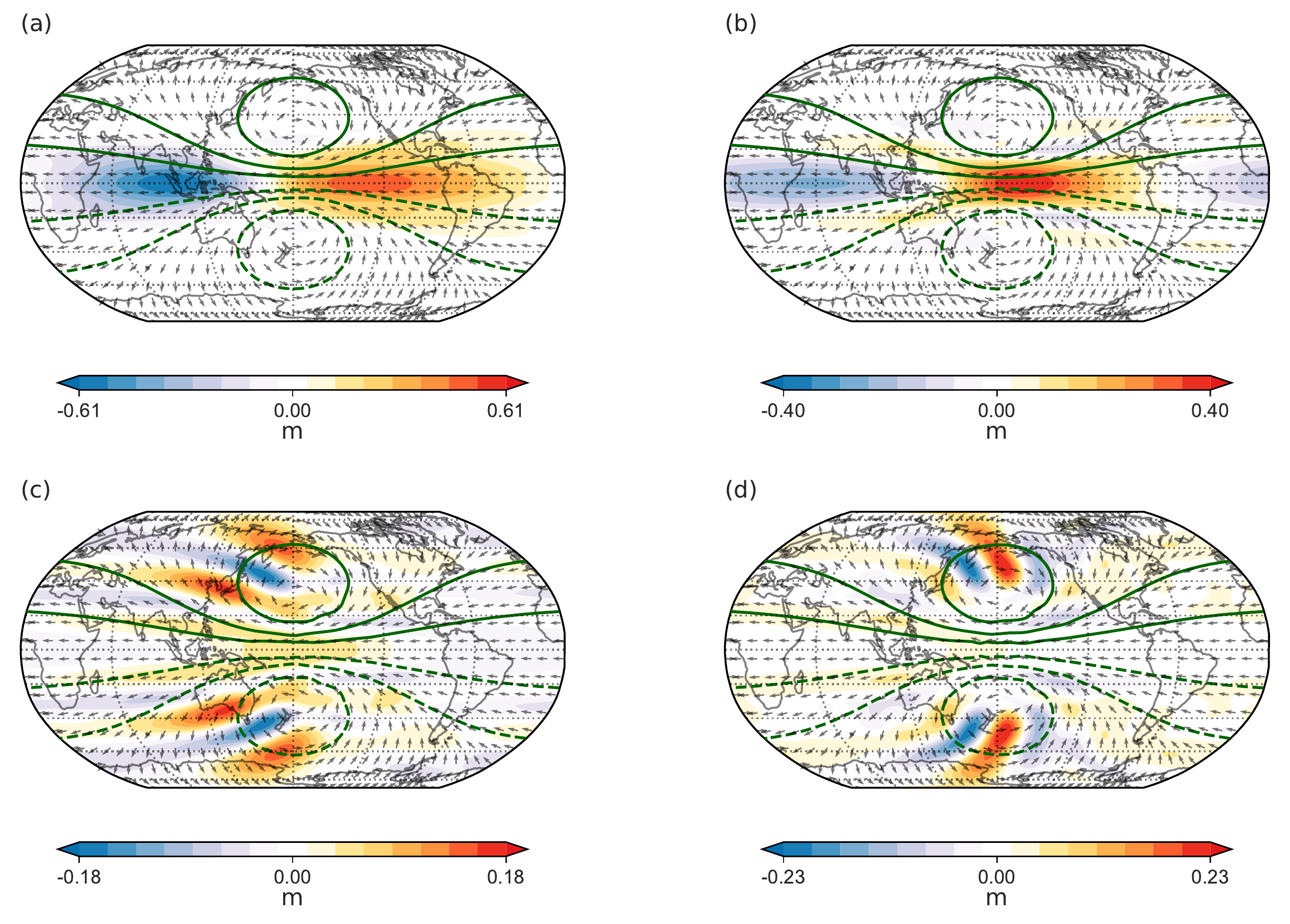}
    \caption{Eastward component of height anomalies at (a) Day25, (b) Day 50, (c) Day 150 and (d) Day 250. 
    First ten zonal wavenumbers are retained and a 50 day window is used. The moist PV is shown in green contours (solid : positive and dashed : negative values). PV contours are in intervals of $2 \times 10^{-7} m^{-1}s^{-1}$, with zero contour not shown.
    The $q_s$ field and initial Gaussian imbalance are as shown in Figure \ref{fig:gillQsatBothHeight}. The arrows represent the direction of "moist Rossby" waves and points 90$^\circ$ counter clockwise to the local moist PV gradient. }
    \label{fig:gillQsatBothEastFiltHeight}
\end{figure}

\noindent The westward component of height anomalies is again dominated by the large-scale Rossby waves (not shown). Its quadrupole structure is confined to sub-tropics, but has a greater meridional extent as compared to the case where $q_s$ was purely a function of latitude. This is in line with a larger deformation scale (approximately 12$^\circ$) in the present situation. Given that the response in Figure \ref{fig:gillQsatBothHeight} moves out to the midlatitudes, we construct a Ho{\"v}moller plot in the latitude band of 30$^\circ$N to 60$^\circ$N. As seen in Figure \ref{fig:gillQsatBothHovm}, this immediately suggests cohesive eastward movement between 120$^\circ$ to 240$^\circ$ longitudes.
Thus, in contrast to the case when $q_s$ was purely a function of latitude, a feature that stands out is the presence of an eastward moving low-frequency mode. We isolate this eastward component and its evolution with time is shown in Figure \ref{fig:gillQsatBothEastFiltHeight}. 
Quite clearly, the eastward signal begins as a Kelvin wave restricted to the tropics (as in Day 25) and then arcs out to the midlatitudes (Day 50 onwards). In fact, as can be seen on Day 50 in Figure \ref{fig:gillQsatBothEastFiltHeight}, the height anomalies associated with the eastward component begin to spread out to higher latitudes. By Day 250, this signal has curved back towards the tropics from the midlatitudes. Indeed, the midlatitude extension seen in Figure \ref{fig:gillQsatBothHeight} is principally due to the eastward component of the solution.
Similar features are seen when the source position is aligned with, and to the east of, the peak in $q_s$. Judging from Figure \ref{fig:gillQsatBothEastFiltHeight}, in the Northern hemisphere this signal starts propagating in a North-East direction (to the midlatitudes) and then changes to a South-East direction (to the tropics). Further, the velocity anomaly quivers show this off-equatorial response to be primarily rotational in character moving eastward at a speed of about 0.5$^\circ$/day.

\begin{figure}
    \centering
    \includegraphics[width=0.5\textwidth]{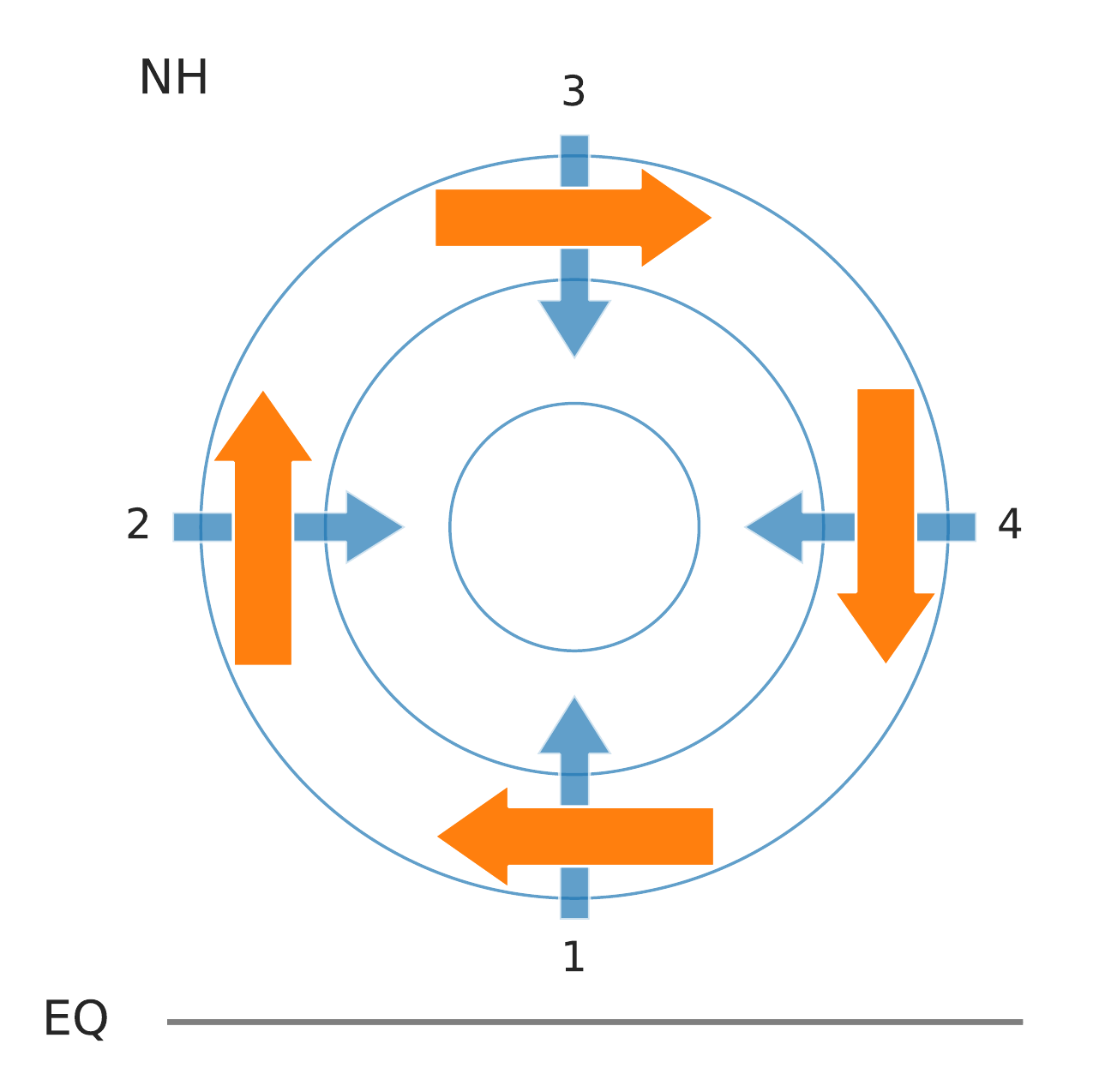}
    \caption{An illustration on the moist PV gradients and eastward propagation. Blue circles are the moist PV contours whose magnitude increases as we go from outer to inner layers. This is similar to the moist PV of Figure \ref{fig:gillQsatBothEastFiltHeight} in the Northern hemisphere and around the dateline. Moist PV gradients and Rossby wave propagation are represented by thin blue and thick orange arrows respectively. }
    \label{fig:PVGradientCartoon}
\end{figure}

\noindent The movement of this eastward component can be explained in terms of moist PV. In the absence of mass and momentum sources (or sinks), Equation \ref{eqn:MoistPV} implies the conservation of moist PV. Similar to the dry Rossby waves which rely on the conservation of PV, here we have an equivalent moist rotational wave (i.e., "moist Rossby" wave) that relies on the conservation of moist PV. 
Further its direction of propagation can be well described by the moist PV gradients
as illustrated in Figure \ref{fig:PVGradientCartoon}. Here, the blue circles are the moist PV contours and its magnitude increases from outer to inner circles. This is similar to the moist PV field in Figure \ref{fig:gillQsatBothEastFiltHeight} around the dateline and in  the Northern hemisphere. Thin blue arrows indicate the direction of the moist PV gradient. Moist Rossby waves propagate to the "west" (90$^\circ$ counter clockwise) of the moist PV gradient as indicated by the thick orange arrows. At position 1,  moist PV increases poleward and as per their dry counterparts, the moist Rossby waves propagate in the westward direction. At position 2, moist PV gradient points east and the moist Rossby waves propagate to the North. The moist PV gradient is "inverted" at position 3 and points equatorward, resulting in an eastward propagating moist Rossby waves. Similarly at position 4 the moist PV gradient is towards the west with equatorward moving moist Rossby waves. In essence, moist PV gradient has a clockwise rotation in the Northern Hemisphere, with the associated moist Rossby waves experiencing a similar rotation but with a 90$^\circ$ phase lag as denoted by the arrows in Figure \ref{fig:gillQsatBothEastFiltHeight}. Thus, when the saturation field was only dependent on latitude, the system could not support a midlatitude excursion and eastward propagation of the moist rotational mode. 

\noindent The ratio of eastward to westward energy hovers around 0.1 to 0.2 at later times (not shown). This indicates that the westward mode has most of the energy, but also suggests that the eastward component does not die away (mainly in the midlatitudes) as was the situation when $q_s$ was only dependent on latitude. 
KE spectra (not shown) indicate the dominance of rotational portion of the flow and the inhibition of energy transfer to smaller scales. But this inhibition is not as strong as in the purely meridionally varying saturation profile. 

\subsection{Moist solutions with a realistic saturation profile}

The responses studied so far have been in the presence of idealized background saturation fields. To see if the intuition developed carries over to more realistic scenarios, given the relation between precipitation and column water vapor \citep{MuEm}, we now base the saturation fields on climatological mean precipitable water vapor for the months of January and July obtained from NCEP-NCAR reanalysis data. Further, any connection to real world phenomena will also be most likely apparent here as the summer and winter responses of the atmosphere are indeed guided by these spatially inhomogeneous saturation environments. In fact, it has been noted that seasonal variations in the MJO are dependent on the mean moisture field \citep{jiang}. Note that, to have a smooth $q_s$ field only the first five spherical harmonics are considered. 

\begin{figure}
    \centering
    \includegraphics[width=0.5\textwidth]{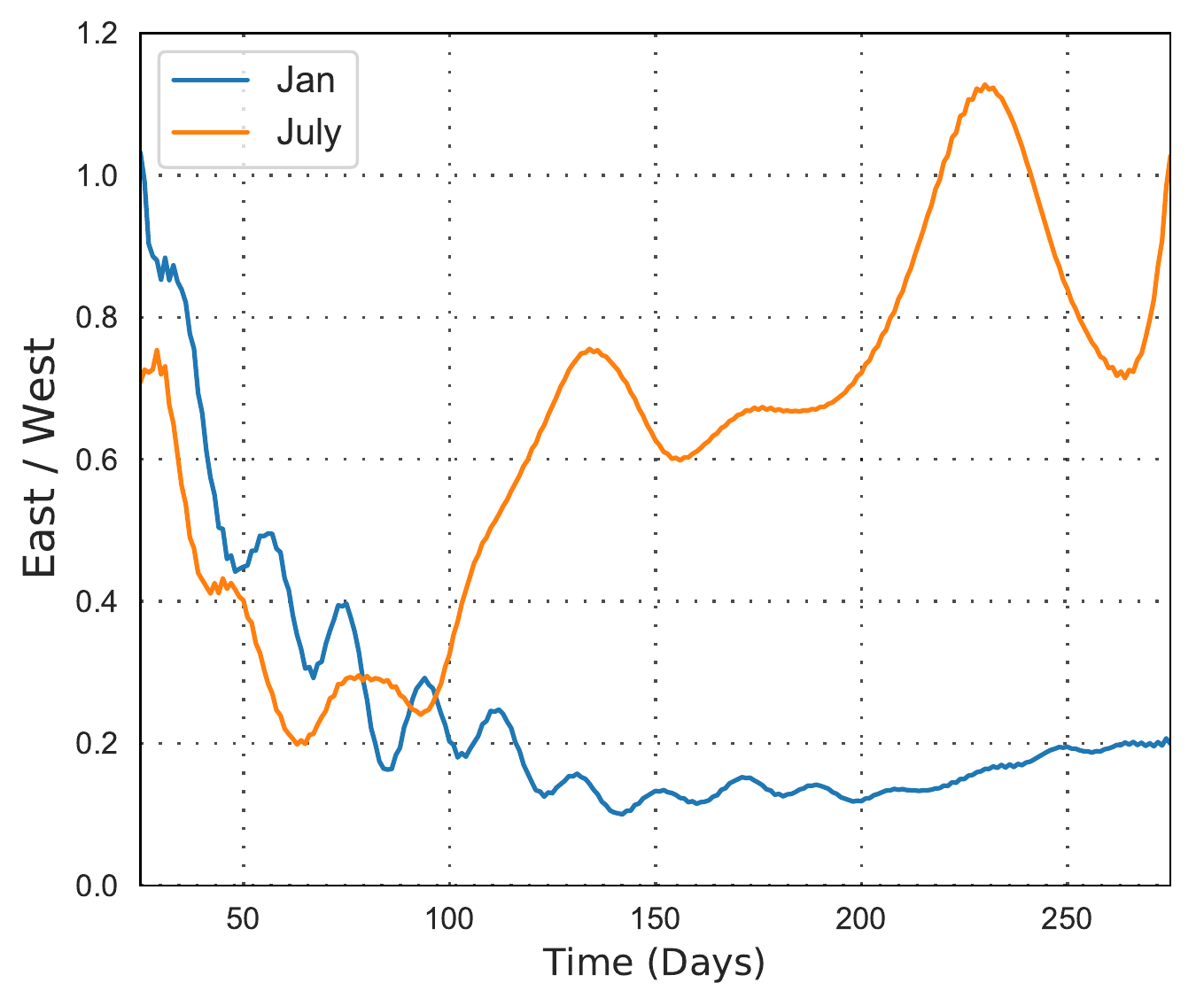}
    \caption{Evolution of eastward to westward ratio of energy (KE+PE) with time. Here, $q_s$ is derived from the  climatological mean precipitable water vapor for the months of January and July obtained from NCEP-NCAR reanalysis data. }
    \label{fig:gillQsatActualEnergyTime}
\end{figure}

\noindent For both January and July cases we obtain robust eastward and westward moving components. This is evident in the ratio of eastward to westward energy shown in Figure \ref{fig:gillQsatActualEnergyTime}; in fact, in the boreal summer, there is almost an equal amount of energy in the two components. The wavenumber-frequency diagrams for the two seasons constructed from data within 30$^\circ$ of the equator (see Figure S1 in supporting information) indicate the energy of the initial response to be concentrated along the dispersion curves of Rossby (westward) and Kelvin (eastward) waves whose equivalent depth matches with that estimated from the linear rapid condensation limit. 
At later times, especially for the July case, there is a considerable energy in slower (lower frequency) modes at higher wavenumbers.

\begin{figure}
    \centering
    \includegraphics[width=\textwidth]{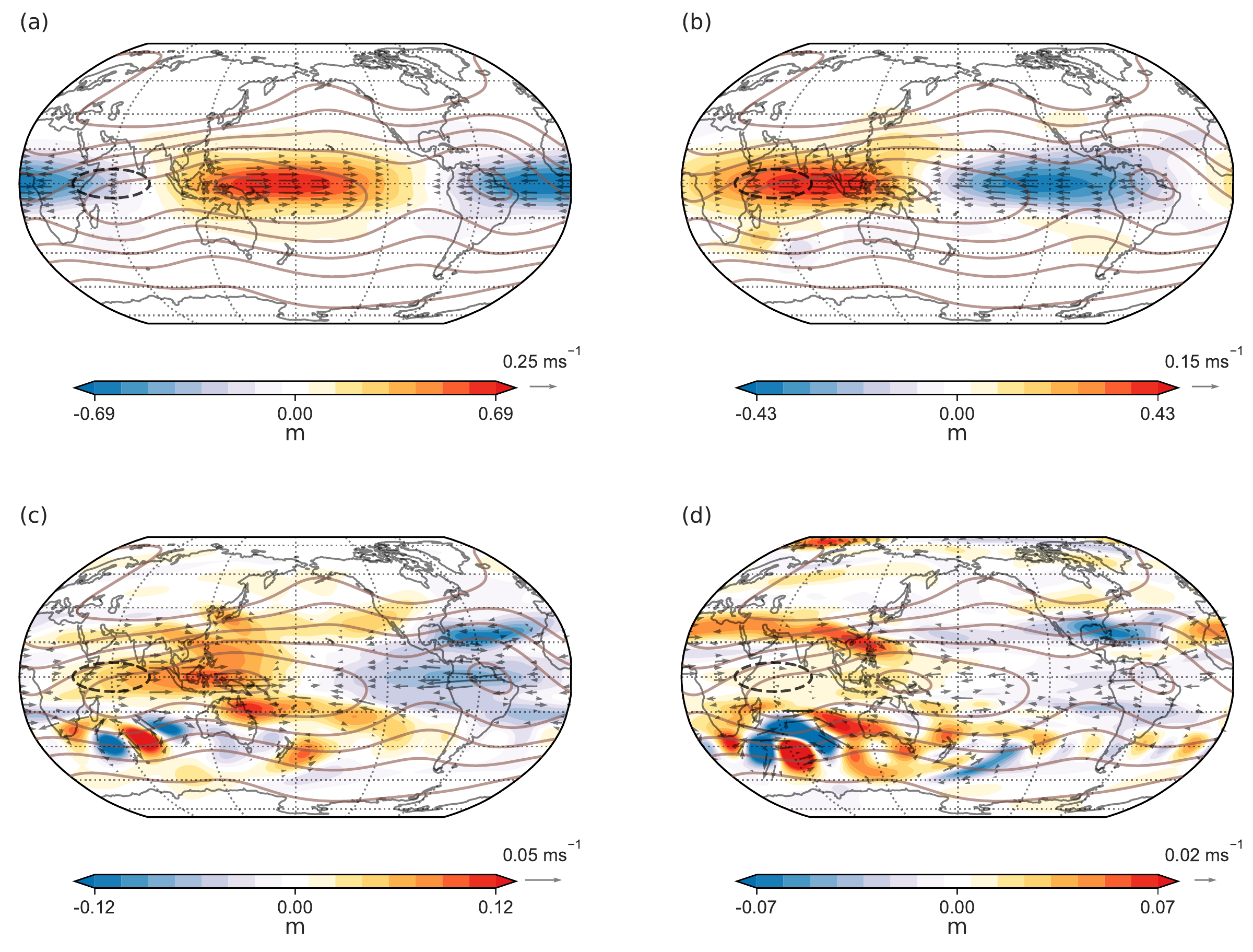}
    \caption{Eastward component of height anomalies at (a) Day 25, (b) Day 50, (c) Day 150 and (d) Day 250 for the moist case during the boreal winter. First ten zonal wavenumbers are retained and a 50 day window is used. The system is initially forced by a Gaussian mass sink (black dashed contours). Here, $q_s$ is derived from the January climatology of precipitable water vapor obtained from NCEP-NCAR reanalysis data (shown in brown contours). The saturation contours are in intervals of 0.008 m, with zero contour not shown.}
    \label{fig:gillQsatActualJanEastFiltHeight}
\end{figure}

\begin{figure}
    \centering
    \includegraphics[width=\textwidth]{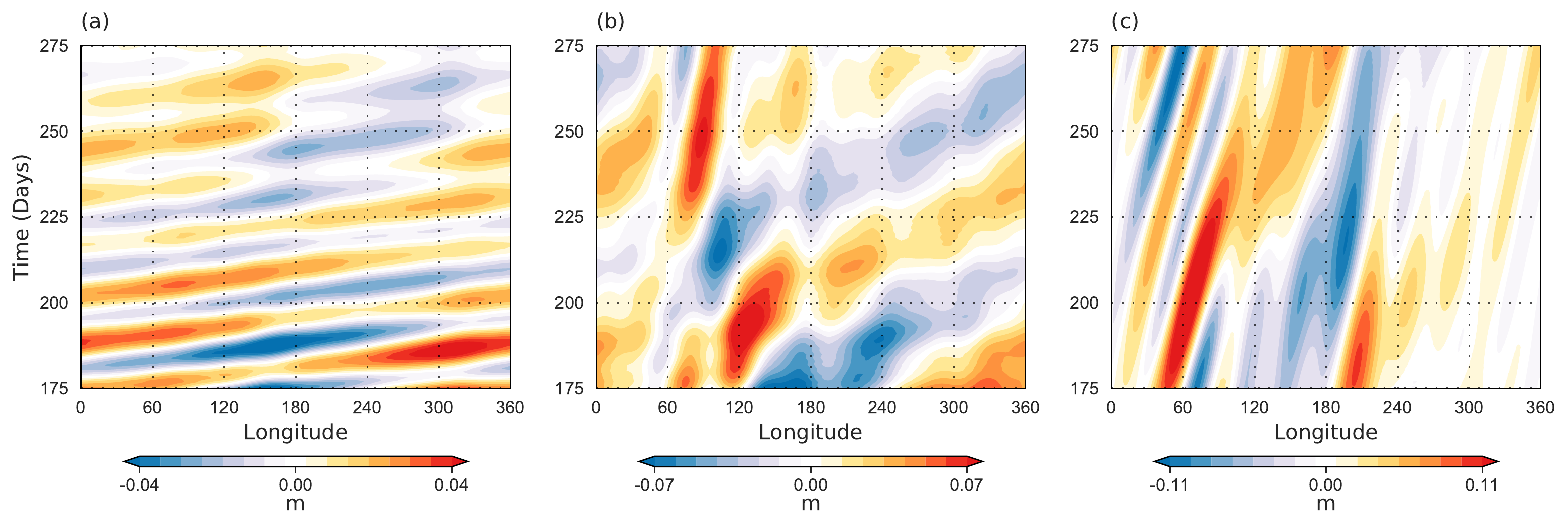}
    \caption{Ho{\"v}moller plot  of eastward component of height anomalies for the moist case during the boreal winter. The system is initially forced by a Gaussian mass sink and $q_s$ is derived from the January climatology of precipitable water vapor obtained from NCEP-NCAR reanalysis data. The three panels are for (a) tropics (0$^\circ$S - 10$^\circ$S), (b) sub-tropics (15$^\circ$S - 25$^\circ$S) and (c) mid-latitudes (30$^\circ$S - 60$^\circ$S).}
    \label{fig:gillQsatJanHovmEast}
\end{figure}

\noindent Focusing on the eastward propagating response, this component's height and velocity anomalies for the January case are presented in Figure \ref{fig:gillQsatActualJanEastFiltHeight}. As expected, a clear Kelvin wave signature is seen early (Day 25 and 50) in the simulation. This, at later stages, changes to a quadrupole structure (Figure \ref{fig:gillQsatActualJanEastFiltHeight}; Day 150 and 250) in the tropical belt. In fact, large-scale height anomalies on the equator persist till Day 150, but are largely absent by Day 250 when this field is dominated by off-equatorial extrema. Note that the quadrupole is restricted to very large scales and its magnitude is smaller than the initial Kelvin wave.  
At Day 250 this consists of positive anomalies over Asia and Australia that extend westward up to Africa and is complemented by negative anomalies over the American continents. The velocity anomalies suggests that this quadrupole is rotational in character. 
Mid-latitude eastward moving anomalies (not seen in the wavenumber-frequency diagrams as they are restricted to 30$^\circ$ N/S) also form near the Southern Africa and Australia in the Southern Hemisphere and to a weaker extent over Northern Asia and the Pacific in the Northern Hemisphere. 
Following the schematic in Figure \ref{fig:PVGradientCartoon}, the moist PV gradient (not shown) suggests south-eastward and north-eastward movement over south-eastern Africa and the western Pacific, respectively. This agrees with the strongest anomalies seen in Figure \ref{fig:gillQsatActualJanEastFiltHeight} on Days 150 and 250. 
The changing nature of this eastward response can also be seen via the Ho{\"v}moller diagrams in Figure \ref{fig:gillQsatJanHovmEast}. Specifically, near the equator (0$^\circ$ to 10$^\circ$ S), we observe Kelvin waves that decay and also break into smaller scale features with time. In the subtropics (15$^\circ$ to 25$^\circ$ S), the quadrupole structure emerges and its propagation is slower than the Kelvin wave and its speed is also dependent on longitude. In particular, it is worth noting that this structure is stronger and moves comparatively more slowly in the eastern hemisphere (especially from 60$^\circ$ to 120$^\circ$ longitude) where there is a higher precipitable water available for moist coupling. In the midlatitudes (30$^\circ$ to 60$^\circ$ S), an even slower moving signal is apparent and is largely confined to the eastern hemisphere. While the eastward component changes quite strikingly with time, the westward mode remains fairly consistent with the linear Rossby dispersion curve and consists of cyclonic and anticyclonic eddies that straddle the equator (Figure S2 in supporting information).  

\begin{figure}
    \centering
    \includegraphics[width=\textwidth]{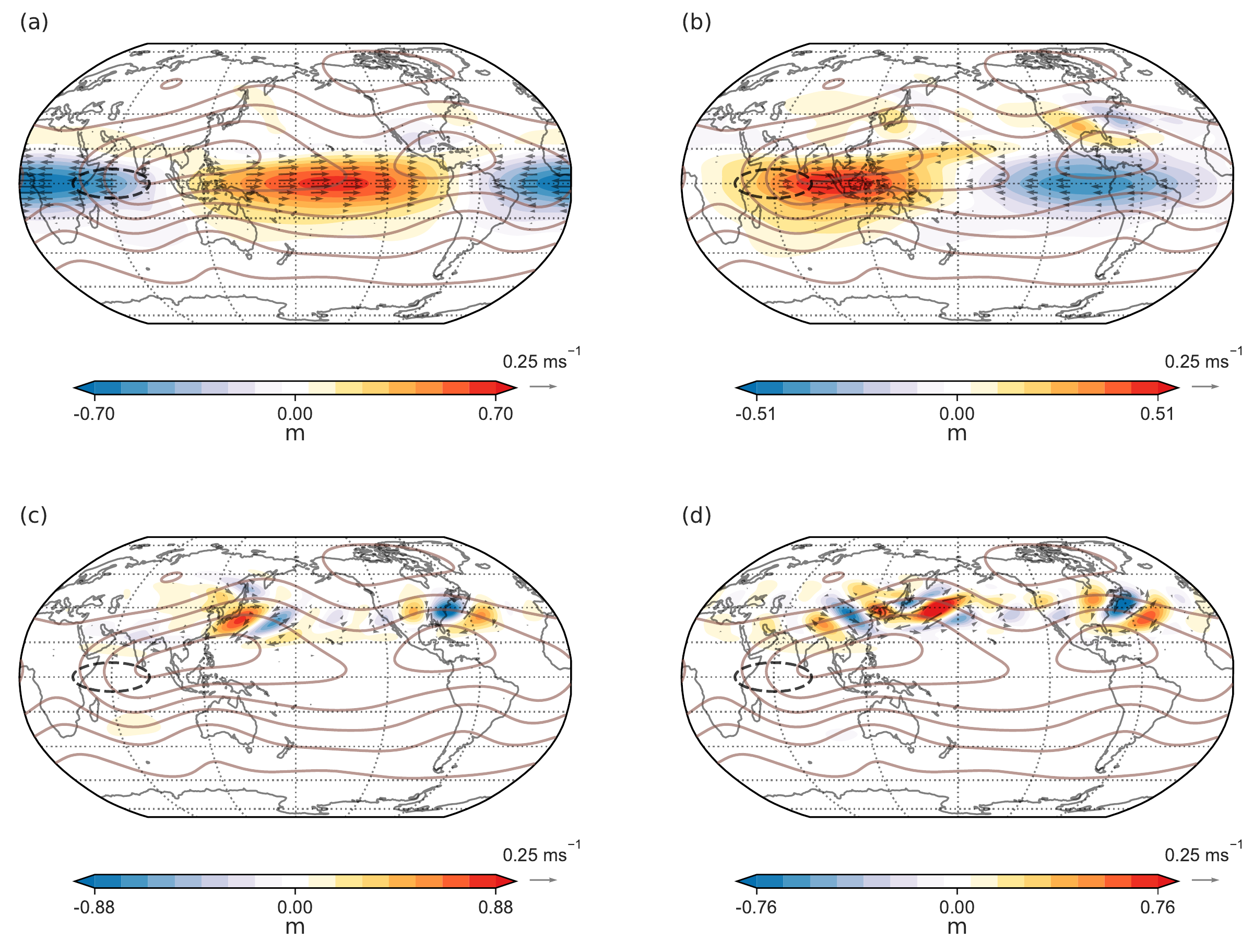}
    \caption{Eastward component of height anomalies at (a) Day 25, (b) Day 50, (c) Day 150 and (d) Day 250 for the moist case during the boreal summer. First ten zonal wavenumbers are retained and a 50 day window is used. The system is initially forced by a Gaussian mass sink (black dashed contours). Here, $q_s$ is derived from the July climatology of precipitable water vapor obtained from NCEP-NCAR reanalysis data (shown in brown contours). The saturation contours are in intervals of 0.008 m, with zero contour not shown.}
    \label{fig:gillQsatActualJulyEastFiltHeight}
\end{figure}

\noindent In contrast, the July saturation field yields a response, shown in Figure \ref{fig:gillQsatActualJulyEastFiltHeight}, that begins as a Kelvin wave --- much like in January --- and then appears as a predominant eastward moving wave train originating in the subtropics, moving over the Tibetan plateau region into the midlatitudes up to North America. As this signal extends into the midlatitudes it is not completely captured in the wavenumber-frequency diagram (Figure S1 in supporting information), but the late time features clearly suggest a long time period mode that is spread out across multiple spatial scales. In fact, during the boreal summer the longer time features are almost exclusively restricted to the Northern Hemisphere. Note that the westward response (Figure S3 in supporting information) is present in both hemispheres. The eastward propagation is guided by the moist PV mechanism outlined in Figure \ref{fig:PVGradientCartoon}, as is evident from the contours of precipitable water and the position of height anomalies. Indeed, in the boreal summer, precipitable water has a maximum off the equator in the Bay of Bengal \citep{AdamesMing}, and this guides the atmospheric response towards the subtropics over the Indian landmass. Also it is worth noting that the strength of eastward anomalies remain relatively steady with time. 
Once again, the velocity anomalies along with the height anomaly fields suggest that this eastward signal is predominantly rotational in character. In all, Figure \ref{fig:gillQsatActualJulyEastFiltHeight} suggests a teleconnection from a heat source in the equatorial western Indian Ocean to North America. Thus, in both seasons, the initially generated Kelvin wave decays, and to some extent, emergent eastward propagating modes appear to be guided by moist PV gradients.

\begin{figure}
    \centering
    \includegraphics[width=\textwidth]{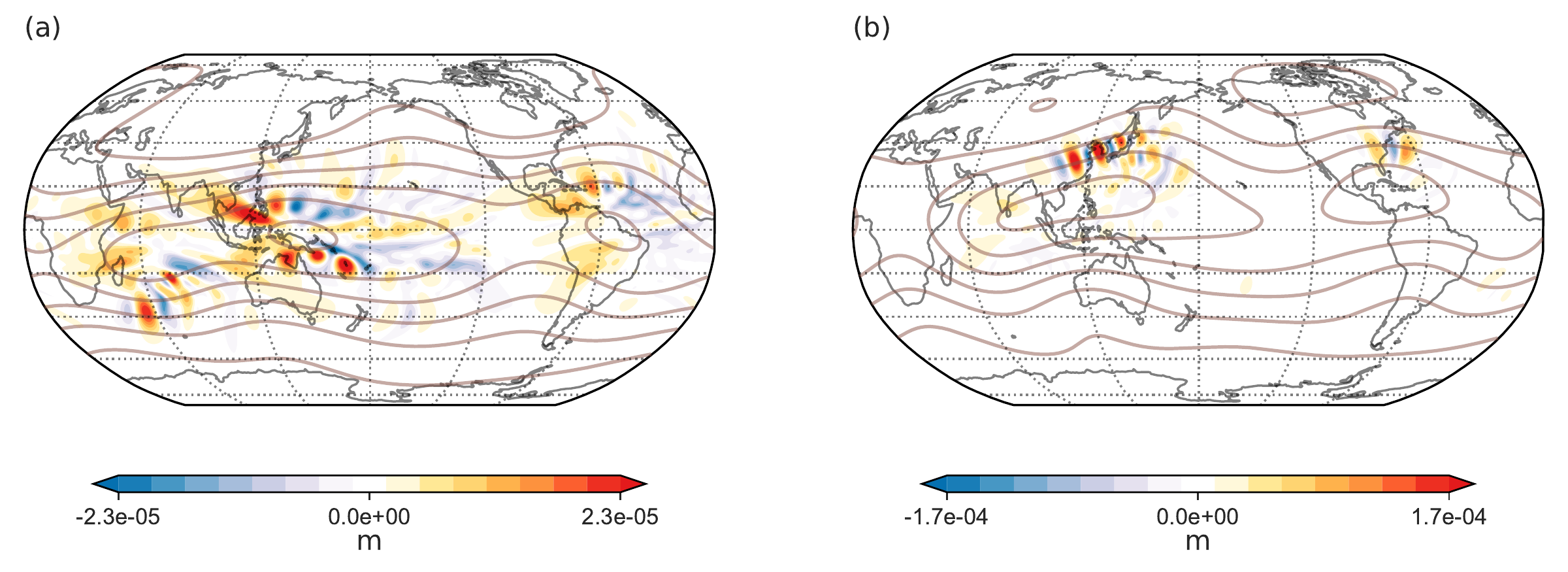}
    \caption{Moisture anomaly (deviation from the saturation state) at Day 250 for (a) January and (b) July saturation profiles (shown in brown contours), respectively.  Positive (negative) values indicate the region of precipitation (evaporation). The saturation contours are in intervals of 0.008 m, with zero contour not shown. The system is initially forced by a Gaussian height imbalance as shown in Figure \ref{fig:gillQsatActualJulyEastFiltHeight}.}
    \label{fig:gillQsatActualQp}
\end{figure}

\noindent The condensation and evaporation fields for both January and July, are shown in Figure \ref{fig:gillQsatActualQp}.  For July, a strong North-East oriented wave train is seen that spans the Indian sub-continent to the western Pacific. For an observer located in India this appears as an oscillation (wet and dry) with an approximate time period of 40 to 50 days. Note that as the wave train progresses into the midlatitudes the condensation and evaporation anomalies are absent (or weak) and reappear as the signal curves back towards the equator over the southern tip of North America. For January, consistent with the background precipitable water field, moisture anomalies are mainly seen in the eastern hemisphere, and as noted in Figure \ref{fig:gillQsatJanHovmEast}, the signal propagates slowly in this region. 
Comparing Figure \ref{fig:gillQsatActualQp} with the Day 250 height anomalies in Figures \ref{fig:gillQsatActualJanEastFiltHeight} and \ref{fig:gillQsatActualJulyEastFiltHeight}, we notice that moisture anomalies are tied to the eastward propagating response in each season. 
Further, in both January and July, even though the initially imposed convergence was on the equator, at later times both the westward and eastward rotational response conspire to create off-equatorial maxima in this field (not shown), and are aligned with regions of condensation.

\section{Discussion and Conclusions}

\noindent In this work we have examined the response of a nonlinear moist spherical shallow water system to imbalances in the tropics. In particular, our aim has been to highlight the dependence of the response on differing background saturation environments. In fact, much like the change in the response to tropical heating on inclusion of mean flows \citep{LauLim,Sardesh}, given that saturation field in reality is horizontally inhomogeneous \citep{Sukhatme1}, these results are expected to serve as a useful guide in interpreting propagating disturbances from tropical forcing. 

\noindent The moist solution is first explored with a purely meridional dependency in the saturation field. 
Early on we see the emergence of Rossby and Kelvin waves, both of which were confined to the subtropics. These waves move slower than their dry reference counterparts, and their speed is consistent with a reduced equivalent depth that matches well with the predictions from linear rapid condensation estimates. Quite strikingly, in contrast to the dry simulations, the eastward component decays and at long times only the westward moving Rossby gyres remains, i.e., the initial imbalance adjusts to a westward propagating Rossby quadrupole. Indeed, initially formed Kelvin waves decay rapidly in all the moist simulations. It appears that the primary driving of moisture anomalies by Rossby modes lead to their misalignment with equatorial Kelvin wave convergence zones.
Further, the off-equatorial position of condensation maxima between adjoining gyres is in accord with outgoing longwave radiation anomalies observed in convectively coupled equatorial Rossby waves \citep{WKW}. At longer times, KE spectra suggest no significant transfer of energy to the smaller scales, and in contrast to the dry run, KE is dominated by the rotational portion of the flow. 

\noindent When the background saturation field is allowed to vary both in the latitudinal and longitudinal directions, in addition to westward propagating modes, a distinct low frequency eastward moving component is observed at long times. This component arcs out to the midlatitudes (to the west of saturation maxima), propagates in the mid-latitudes and then curves back to the tropics (east of the saturation maxima). Further, this slow moving eastward mode is rotational in character and relies on the conservation of moist PV. Its meridional and eastward propagation can be understood in terms of the moist PV gradient, which in turn is affected by the background  saturation field. In particular, the moist PV gradient has a clockwise rotation in the Northern Hemisphere, with the associated moist Rossby waves experiencing a similar rotation but with a 90$^\circ$ phase lag.

\noindent As mentioned, previous investigations with moist shallow water systems also reported the existence of slow eastward propagating modes. In particular, when strongly coupled to precipitation, \citet{Yama} noted a spontaneous eastward movement of an imposed region of heating, \citet{Solod} incorporated wind evaporation feedback and found mixed Rossby-gravity waves to play an important role in generating a forced eastward response and \citet{Yang} attributed intraseasonal eastward activity to be due to an interference of inertia-gravity waves. More recently, eastward modes have been noted by \citet{wang2016} who highlighted the role of boundary layer convergence that preceded convection and moisture that coupled the emergent Kelvin and Rossby waves, \citet{Vallis} who found preferential convective aggregation on the eastern side of the system and \citet{RoZe2,RoZe1} who found these modes to arise from special initial conditions and more generally during adjustment to large amplitude pressure anomalies on an equatorial $\beta$-plane. Further, linear analyses with moisture gradients showed the presence of unstable eastward modes in certain parameter regimes \citep{Sobel,Sukhatme2,joyQG}. Indeed, the eastward propagation observed in this work adds to this body of literature but also differs in some important ways, namely, there are no immediate thresholds for eastward activity and furthermore, the physical basis of the modes appears to be the conservation of moist PV.

\noindent Finally, more realistic scenarios with saturation environments following climatological mean precipitable water vapor estimated from reanalysis were considered. The initial response in both seasons consisted of Rossby and Kelvin waves that aligned with linear dispersion curves estimated form the rapid condensation limit. At late times, the westward component still had a peak coinciding with circumnavigating Rossby waves, but there was a sign of energy in smaller scales and at longer time periods. The late time eastward propagating signal on the other hand was markedly different in the two seasons. Specifically, in the boreal winter, this component changed from a tropically confined Kelvin wave to a large-scale quadrupole structure in the subtropics. Along with this rotational quadrupole, the response also extended out into the midlatitudes. 
Further, given the variation in the moist background, the speed of propagation of the eastward quadrupole depended on longitude. 
In contrast, during the boreal summer, precipitable water has a maximum off the equator in the Northern Hemisphere (over the Bay of Bengal) which results in an inverted moist PV gradient (pointing towards the equator). This supported a predominantly eastward moving signal originating in the subtropics. Further, the later time eastward component was spread out across multiple spatial scales, but restricted to large time periods. Associated with this, the precipitation field showed a strong North-East oriented wave train from the Indian sub-continent to the west Pacific. For an observer fixed in this region, this results in wet and dry oscillations that have a time period of about 40 to 50 days. Also, notably, in the boreal summer, the long time eastward response was confined to the Northern Hemisphere and brings out a teleconnection from the equatorial Indian Ocean to the North American region. It is appropriate at this stage to 
keep in mind that these are initial value problems and the exact nature of the solution (such as its phase), especially at long times, will likely depend on the parameters used in the model as well as the form of the initial conditions. Although preliminary work indicates that the qualitative nature of the long time solution remains valid for perturbations to the initial conditions, model parameters and resolution, a detailed study involving ensembles is in progress to benchmark and lend more confidence to the robustness of these solutions. 

\noindent Taken together, the simple moist shallow water model we have considered clearly brings out the changing atmospheric response to tropical forcing with different background saturation environments. Overall we see a picture where the initial moist response is similar to the dry problem (of course, the waves produced propagate slower), but at late times, the emergent solution is very different in character. Indeed, 
from a conceptual perspective, arguably, the two most striking results obtained are (a) The change in the solution when the saturation field goes from being only dependent on latitude to being allowed to vary in both the latitudinal and longitudinal directions. Specifically, the long time solution was restricted to a westward Rossby quadrupole in the former, while the latter shows an additional eastward component --- a "moist Rossby" wave --- that extends out into the midlatitudes. (b) The differences in the eastward component during the boreal winter and summer where
we observe a subtropical quadrupole (with accompanying 
midlatitude extensions) and a northern hemispheric wavetrain from the tropics to the midlatitudes, respectively. Indeed, these perhaps remind one of the eastward moving MJO \citep{AW} with moist convection and a slower speed of propagation in the eastern hemisphere \citep{HS,SK}, and the northward propagating summer intraseasonal oscillation \citep{biso}, in the winter and summer seasons, respectively.
While these are fairly idealized experiments, and it would be beyond their scope to make a direct connection to either the MJO or the boreal summer intraseasonal oscillation,
the results serve as a guide to anticipated large scale atmospheric responses to tropical forcing in the presence of inhomogenous saturation fields in each season. In particular, apart from circumnavigating westward moving Rossby waves, eastward propagating modes are naturally produced when the saturation field depends on latitude and longitude, thus showing that such intraseasonal activity is intrinsic to the shallow water system with interactive moisture. Moreover, as a generalization of dry Rossby waves, these rotational eastward modes appear to be a consequence of the conservation of moist potential vorticity.

\section*{Acknowledgements}
We thank Aditya Baksi for related numerical experiments on the response of the dry shallow water system to tropical forcing. Comments by all the three reviewers and associate editor are gratefully acknowledged as they helped improve the manuscript considerably.

\bibliographystyle{apalike}
\bibliography{ref.bib}

\end{document}


\maketitle

\begin{figure}
    \centering
    \includegraphics[width=\textwidth]{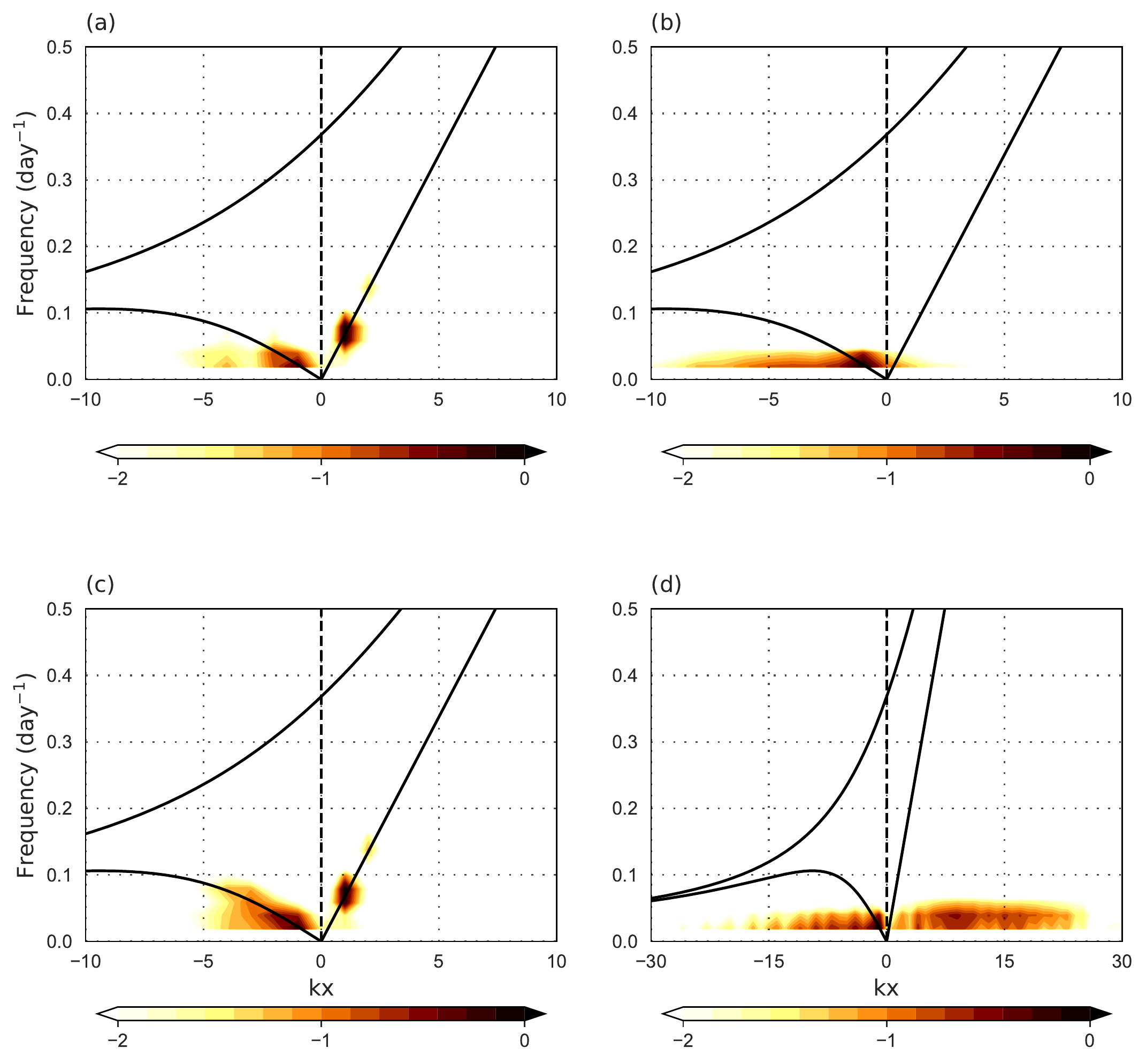}
    \caption{Wavenumber frequency plot of height anomaly field for the realistic $q_s$ derived from the January and July climatology of precipitable water vapor obtained from NCEP-NCAR reanalysis data. Early (Day 1 - 50) and late (Day 251-300) periods for January case are in panels (a) and (b) while for the July case is plotted in (c) and (d) respectively. The system is initially forced by a Gaussian mass sink. The latitudinal band used is from 30$^\circ$N to 30$^\circ$S. The power is normalized and plotted in log scale. Dispersion curves are plotted for an equivalent depth of 100 m.}
    \label{fig:gillQsatActualWK}
\end{figure}

\begin{figure}
    \centering
    \includegraphics[width=\textwidth]{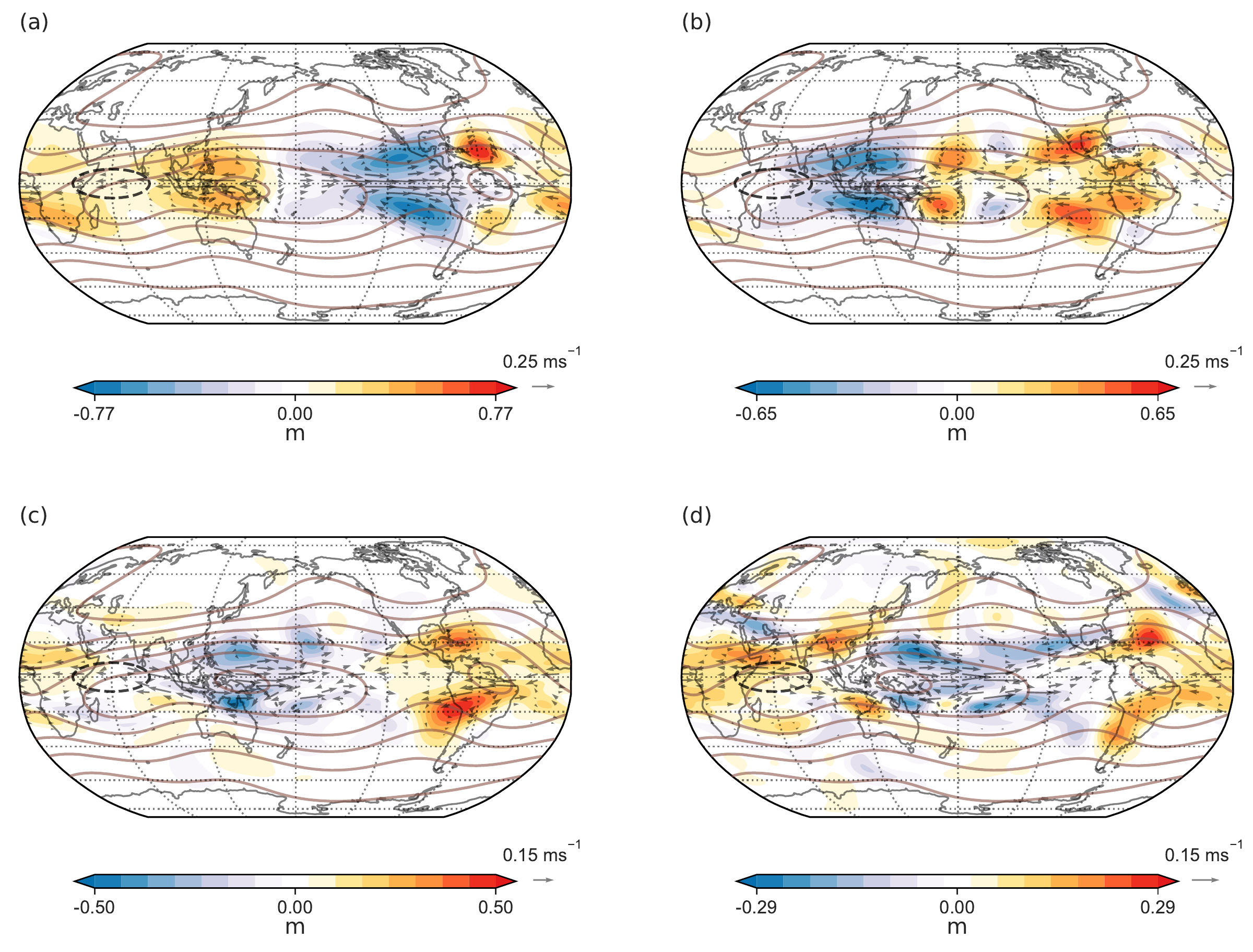}
    \caption{Westward component of height anomalies at (a) Day 25, (b) Day 50, (c) Day 150 and (d) Day 250 for the moist case during the boreal winter. First ten zonal wavenumbers are retained and a 50 day window is used. The system is initially forced by a Gaussian mass sink (black dashed contours). Here, $q_s$ is derived from the January climatology of precipitable water vapor obtained from NCEP-NCAR reanalysis data (shown in brown contours). The saturation contours are in intervals of 0.008 m, with zero contour not shown}
    \label{fig:gillQsatActualJanWestFiltHeight}
\end{figure}

\begin{figure}
    \centering
    \includegraphics[width=\textwidth]{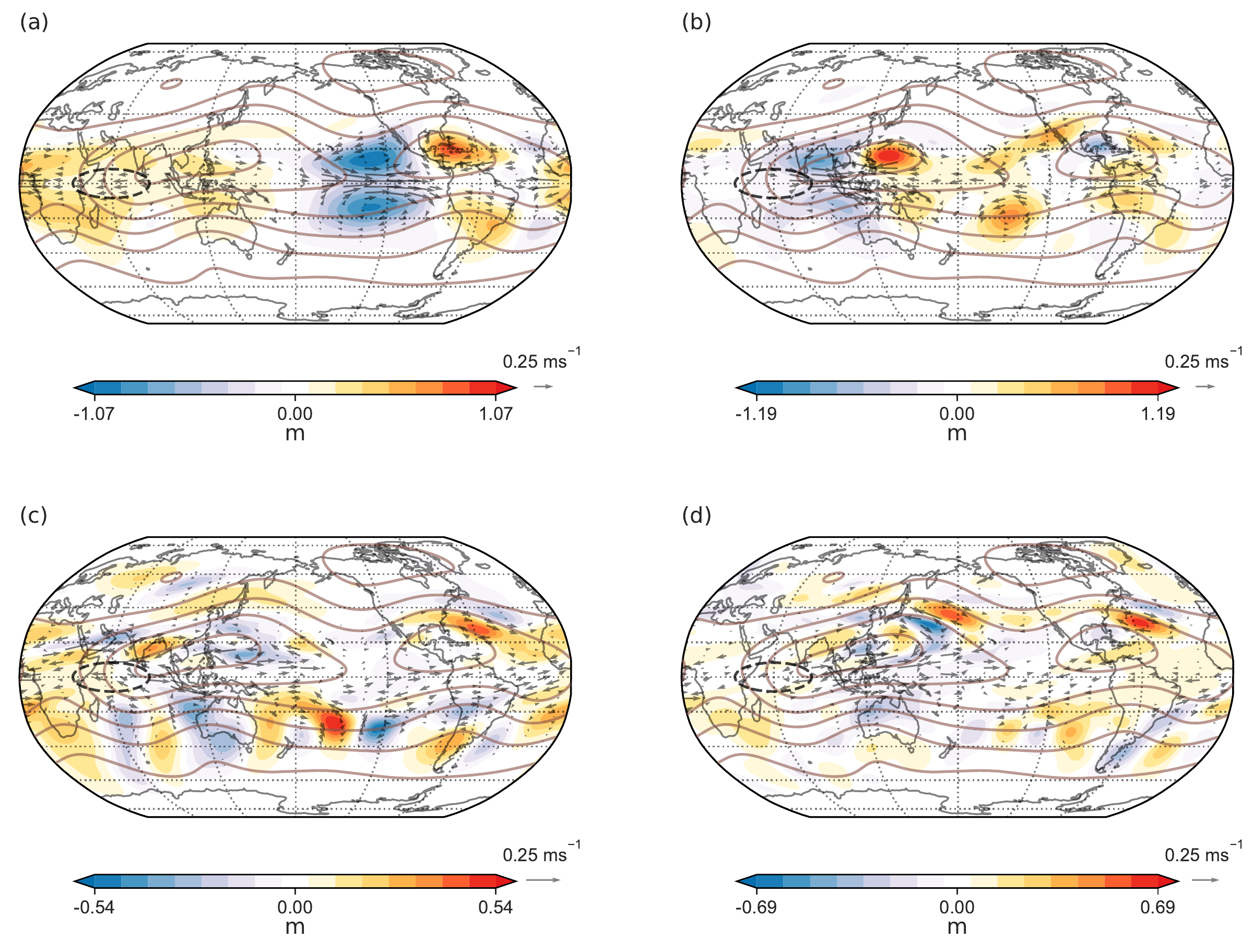}
    \caption{Westward component of height anomalies at (a) Day 25, (b) Day 50, (c) Day 150 and (d) Day 250 for the moist case during the boreal summer. First ten zonal wavenumbers are retained and a 50 day window is used. The system is initially forced by a Gaussian mass sink (black dashed contours). Here, $q_s$ is derived from the July climatology of precipitable water vapor obtained from NCEP-NCAR reanalysis data (shown in brown contours). The saturation contours are in intervals of 0.008 m, with zero contour not shown.}
    \label{fig:gillQsatActualJulyWestFiltHeight}
\end{figure}